\documentclass[aps,pra,amsmath,amssymb,bibnotes]{revtex4-2}

\usepackage{amsmath}
\usepackage{mathtools}
\usepackage[english]{babel}
\usepackage[utf8]{inputenc}
\usepackage{hyperref}
\usepackage{textcomp} 
\usepackage{physics}
\usepackage{float}
\usepackage{dsfont}
\usepackage{enumitem}
\usepackage{MnSymbol}
\usepackage{comment}
\usepackage{color}
\usepackage{ulem}
\usepackage{natbib}

\begin{document}

\title{Fluctuations of the energy density and intensity for arbitrary objects in an arbitrary environment}
	
\date{\today}
	
\author{Florian Herz$^{*}$}
\affiliation{Laboratoire Charles Fabry, UMR 8501, Institut d'Optique, CNRS, Université Paris-Saclay, 2 Avenue Augustin Fresnel, 91127 Palaiseau Cedex, France}
\email{florian.herz@institutoptique.fr} 
	
\begin{abstract}
I apply the scattering approach within the framework of macroscopic quantum electrodynamics to derive the variances and mean values of the energy density and intensity for a system of an arbitrary object in an arbitrary environment. To evaluate the temporal bunching character of the energy density and intensity, I determine the ratio of their variances with respect to their mean values. I explicitly evaluate these ratios for the cases of vacuum, a half-space in vacuum, and a sphere in vacuum. Eventually, I extend the applicability of this theory to the case of more than one arbitrary object, independent of the geometrical shapes and materials. 
\end{abstract}
\maketitle

\section{Introduction}

In most works on near-field thermal radiation, theory and experiment focus on the analysis of coherence properties of the first order, e.g.\ the heat flux or mean Poynting vector. In contrast, higher order coherence properties for thermal near-field radiation, for instance, the variance of the energy density or the heat flux are only scarcely investigated. From an experimental point of view, the variance or fluctuations around the mean values can only be monitored with improved ultra fast measurement methods. In a theoretical description, this demands the additional assumption of the Gaussian property for thermal radiation in the near-field regime to be able to evaluate the corresponding correlation functions within fluctuational electrodynamics. This property was used to calculate the variance of the Casimir-Lifshitz force \cite{Barton1991, Kogan2005, Messina2007} and the vacuum friction \cite{vac_fric} in the near-field. In recent years, also the fluctuations of thermal quantities moved into the focus of interest, especially when evaluating their impact on experiments, in which the spectral information is lost while measuring the heat currents \cite{wise}, for instance. There are also works on the variance of the mean Poynting vector between two planar media \cite{rhe_flucs}. A very important application of theses higher order correlation functions are Green-Kubo relations which connect the linear transport coefficients of a system out of thermal equilibrium with the equilibrium fluctuations of the corresponding quantity \cite{Golyk, Herz2}. 

The first order spatial coherence property of thermal radiation is well studied in the far- and near-field regime. For instance, for half-spaces it was shown that the coherence length strongly depends on the chosen material. If it supports surface waves, the coherence length can be much larger than the well-known $\lambda/2$ of black-body radiation but if it does not support them, the coherence lengths can be much shorter \cite{spcoh1}. This was later validated by discussing the contributions of surface waves, skin-layer currents, and small-scale polarization fluctuations to the cross-spectral density tensor \cite{spcoh2} as well as by analyzing the energy density with respect to surface waves \cite{shchegrov}. For periodically micro-structured SiC and photonic crystals this can be exploited to confine the emission angles to build an infrared antenna \cite{spcoh3, Laroche2005, spcoh4}. 

For thermal radiation, the expectation value defined by Glauber \cite{glauber} has to be evaluated by using the density matrix formalism because it is a mixed state due to the broad range of frequencies involved. Then, these expectation values can be treated by macroscopic quantum electrodynamics (MQED) formalism introduced by Scheel and Buhmann \cite{mQED}. For some dielectric materials like SiC, the near-field spectrum becomes quasi-monochromatic due to the resonance at the surface phonon polariton (SPhP) frequency. Such a change of the spectrum from broadband in the far-field to quasi-monochromatic in the near-field makes it interesting to investigate the bunching property of the thermal near-field radiation. By employing the scattering approach introduced by Rahi et al.\ \cite{Rahi} and Krüger et al.\ \cite{KruegerEtAlTraceFormulas2012}, the correlation functions necessary to study second order coherence can even be generalized to basis independent expressions. 

In the following, I will derive the mean values and variances of the intensity and the energy density for a system of an arbitrary object in an arbitrary environment. Subsequently, I will compute the degree of coherence which I use to investigate the bunching character of three special systems -- vacuum, a substrate in vacuum, and a sphere in vacuum. Eventually, I also extend this theory to a system of more than one arbitrary object.

\section{Theoretical framework}

In classical electrodynamics the energy density $u$ is given by
\begin{align}
u(\mathbf{r}, t) & = \frac{\varepsilon_0}{2} \mathbf{E}^2(\mathbf{r}, t) +  \frac{\mu_0}{2} \mathbf{H}^2(\mathbf{r}, t)
\label{eq:u_time}
\end{align}
with the electric field $\mathbf{E}$, the magnetic field $\mathbf{H}$, the vacuum's permittivity $\varepsilon_0$, and vacuum's permeability $\mu_0$ which are connected by $\mu_0 \varepsilon_0 = 1/c^2$. Since fluctuational electrodynamics treat thermal fluctuations as sources of the electromagnetic fields, the fields and the energy density become fluctuational quantities. Therefore, general mean values are evaluated. Here, to write down the mean value of the energy density, the squared expressions on the right hand side will be replaced by the correlation functions of the considered field
\begin{align}
\Big \llangle u(\mathbf{r}, t) \Big \rrangle & = \frac{\varepsilon_0}{2} \Big \llangle \hat{E}_i(\mathbf{r}, t) \hat{E}_i(\mathbf{r}, t) \Big \rrangle + \frac{\mu_0}{2} \Big \llangle \hat{H}_i(\mathbf{r}, t) \hat{H}_i(\mathbf{r}, t) \Big \rrangle .
\end{align}
Note that I replaced the fields by quantum mechanical operators denoted by the $\hat{\cdot}$ symbol. Here, I will use the symmetrically ordered operators to obtain the energy density and its fluctuations. In addition, let me introduce the positive and negative frequency field operators $\hat{\mathbf{E}}^{\pm}$ defined by 
\begin{align}
\hat{\mathbf{E}} (\mathbf{r}, t) & = \hat{\mathbf{E}}^{+} (\mathbf{r}, t) + \hat{\mathbf{E}}^{-} (\mathbf{r}, t)
\label{eq:E_sep}
\end{align}
which are simply given by
\begin{align}
\hat{\mathbf{E}}^{\pm} (\mathbf{r}, t) & = \int_0^\infty \frac{\mathrm{d} \omega}{2 \pi} \hat{\mathbf{E}} (\mathbf{r}, \pm \omega) e^{\mp \text{i} \omega t} .
\end{align}
Herein, $\hat{\mathbf{E}}^{+}$ describes the annihilation of a photon and $\hat{\mathbf{E}}^{-}$, its Hermitian conjugate, its creation. By using the definition in Eq.~\eqref{eq:E_sep} and taking advantage of the stationarity of the fields, i.e.\ $\llangle \hat{E}^{+}(\omega) \hat{E}^{-}(\omega') \rrangle \propto 2 \pi \delta (\omega - \omega')$, the energy density becomes
\begin{align}
\Big \llangle u(\mathbf{r}, t) \Big \rrangle & = \int_0^\infty \frac{\mathrm{d} \omega}{2 \pi} \Big\{ \frac{\varepsilon_0}{2} \Bigr[ \Big \llangle \hat{E}_i^{+}(\mathbf{r}, \omega) \hat{E}_i^{-}(\mathbf{r}, \omega) \Big \rrangle + \Big \llangle \hat{E}_i^{-}(\mathbf{r}, \omega) \hat{E}_i^{+}(\mathbf{r}, \omega) \Big \rrangle \Bigr] \notag \\
& \quad + \frac{\mu_0}{2} \Bigr[ \Big \llangle \hat{H}_i^{+}(\mathbf{r}, \omega) \hat{H}_i^{-}(\mathbf{r}, \omega) \Big \rrangle + \Big \llangle \hat{H}_i^{-}(\mathbf{r}, \omega) \hat{H}_i^{+}(\mathbf{r}, \omega) \Big \rrangle \Bigr] \Big\} .
\label{eq:u_raw}
\end{align}
Note that index $i$ indicates Einstein summation over the vector components. In quantum mechanics, however, a measuring process, e.g.\ in a photon interference experiment, is not described by symmetrically ordered operators. That is because a photon is annihilated at the detector during the measuring process. This corresponds to the intensity $I$ defined by normally ordered operators
\begin{align}
\Big \llangle I(\mathbf{r}, t) \Big \rrangle & = \Big \llangle \hat{E}_i^{-}(\mathbf{r}, t) \hat{E}_i^{+}(\mathbf{r}, t) \Big \rrangle + \frac{\mu_0}{\varepsilon_0} \Big \llangle \hat{H}_i^{-}(\mathbf{r}, t) \hat{H}_i^{+}(\mathbf{r}, t) \Big \rrangle \notag \\
& = \int_0^\infty \frac{\mathrm{d} \omega}{2 \pi} \Big\{ \Big \llangle \hat{E}_i^{-}(\mathbf{r}, \omega) \hat{E}_i^{+}(\mathbf{r}, \omega) \Big \rrangle + \frac{\mu_0}{\varepsilon_0} \Big \llangle \hat{H}_i^{-}(\mathbf{r}, \omega) \hat{H}_i^{+}(\mathbf{r}, \omega) \Big \rrangle \Big\}.
\end{align}
Note that $\Big \llangle I \Big \rrangle$ follows from $\Big \llangle u \Big \rrangle$ by dropping the first term in each line of Eq.~\eqref{eq:u_raw} and multiplying by $2/\varepsilon_0$.

For the fluctuations of the energy density, one has to evaluate the correlation function of the energy density. This results in a correlation function of four operators, namely
\begin{align}
\Big \llangle u(\mathbf{r}, t) u(\mathbf{r}', t') \Big \rrangle & = \frac{\varepsilon_0^2}{4} \Big \llangle \hat{E}_i(\mathbf{r}, t) \hat{E}_i(\mathbf{r}, t) \hat{E}_j(\mathbf{r}', t') \hat{E}_j(\mathbf{r}', t') \Big \rrangle \notag \\
& \quad + \frac{\mu_0^2}{4} \Big \llangle \hat{H}_i(\mathbf{r}, t) \hat{H}_i(\mathbf{r}, t) \hat{H}_j(\mathbf{r}', t') \hat{H}_j(\mathbf{r}', t') \Big \rrangle \notag \\
& \quad + \frac{\varepsilon_0 \mu_0}{4} \Big \llangle \hat{E}_i(\mathbf{r}, t) \hat{E}_i(\mathbf{r}, t) \hat{H}_j(\mathbf{r}', t') \hat{H}_j(\mathbf{r}', t') \Big \rrangle \notag \\
& \quad + \frac{\varepsilon_0 \mu_0}{4} \Big \llangle \hat{H}_i(\mathbf{r}, t) \hat{H}_i(\mathbf{r}, t) \hat{E}_j(\mathbf{r}', t') \hat{E}_j(\mathbf{r}', t') \Big \rrangle .
\end{align}
This expression can be rewritten in terms of correlation functions of two operators by exploiting the Gaussian property of thermal radiation yielding
\begin{align}
\Big \llangle u(\mathbf{r}, t) u(\mathbf{r}', t') \Big \rrangle & = \Big \llangle u(\mathbf{r}, t) \Big \rrangle \Big \llangle u(\mathbf{r}', t') \Big \rrangle + \frac{\varepsilon_0^2}{2} \Big \llangle \hat{E}_i(\mathbf{r}, t) \hat{E}_j(\mathbf{r}', t') \Big \rrangle \Big \llangle \hat{E}_i(\mathbf{r}, t) \hat{E}_j(\mathbf{r}', t') \Big \rrangle \notag \\
& \quad + \frac{\mu_0^2}{2} \Big \llangle \hat{H}_i(\mathbf{r}, t) \hat{H}_j(\mathbf{r}', t') \Big \rrangle \Big \llangle \hat{H}_i(\mathbf{r}, t) \hat{H}_j(\mathbf{r}', t') \Big \rrangle \notag \\
& \quad + \frac{\varepsilon_0 \mu_0}{2} \Bigl[ \Big \llangle \hat{E}_i(\mathbf{r}, t) \hat{H}_j(\mathbf{r}', t') \Big \rrangle \Big \llangle \hat{E}_i(\mathbf{r}, t) \hat{H}_j(\mathbf{r}', t') \Big \rrangle \notag \\
& \quad + \Big \llangle \hat{H}_i(\mathbf{r}, t) \hat{E}_j(\mathbf{r}', t') \Big \rrangle \Big \llangle \hat{H}_i(\mathbf{r}, t) \hat{E}_j(\mathbf{r}', t') \Big \rrangle \Bigr] .
\label{eq:uu_raw}
\end{align}
Let me now come back to the variance
\begin{align}
\text{Var}_u & = \text{Var}_u (\mathbf{r}, \mathbf{r}', t, t') = \Big \llangle u(\mathbf{r}, t) u(\mathbf{r}', t') \Big \rrangle - \Big \llangle u(\mathbf{r}, t) \Big \rrangle \Big \llangle u(\mathbf{r}', t') \Big \rrangle. 
\end{align}
This quantity can now be evaluated by inserting Eq.~\eqref{eq:uu_raw} and performing a Fourier transform, giving
\begin{align}
\text{Var}_u & = \int_0^\infty \frac{\mathrm{d} \omega}{2 \pi} \int_0^\infty \frac{\mathrm{d} \omega'}{2 \pi} \Bigg\{ \frac{\varepsilon_0^2}{2} \Bigl[ \mathds{C}_\text{EE}^{\pm ij}(\omega) e^{- \text{i} \omega \tau} + \mathds{C}_\text{EE}^{\mp ij}(\omega) e^{\text{i} \omega \tau} \Bigr] \Bigl[ \mathds{C}_\text{EE}^{\pm ji}(\omega') e^{\text{i} \omega' \tau} + \mathds{C}_\text{EE}^{\mp ji}(\omega') e^{-\text{i} \omega' \tau} \Bigr]^{*} \notag \\
& \quad + \frac{\mu_0^2}{2} \Bigl[ \mathds{C}_\text{HH}^{\pm ij}(\omega) e^{- \text{i} \omega \tau} + \mathds{C}_\text{HH}^{\mp ij}(\omega) e^{\text{i} \omega \tau} \Bigr] \Bigl[ \mathds{C}_\text{HH}^{\pm ji}(\omega') e^{\text{i} \omega' \tau} + \mathds{C}_\text{HH}^{\mp ji}(\omega') e^{-\text{i} \omega' \tau} \Bigr]^{*} \notag \\
& \quad + \frac{\varepsilon_0 \mu_0}{2} \Bigr( \Bigl[ \mathds{C}_\text{EH}^{\pm ij}(\omega) e^{- \text{i} \omega \tau} + \mathds{C}_\text{EH}^{\mp ij}(\omega) e^{\text{i} \omega \tau} \Bigr] \Bigl[ \mathds{C}_\text{HE}^{\pm ji}(\omega') e^{\text{i} \omega' \tau} + \mathds{C}_\text{HE}^{\mp ji}(\omega') e^{- \text{i} \omega' \tau} \Bigr]^{*} \notag \\
& \quad + \Bigl[ \mathds{C}_\text{HE}^{\pm ij}(\omega) e^{- \text{i} \omega \tau} + \mathds{C}_\text{HE}^{\mp ij}(\omega) e^{\text{i} \omega \tau} \Bigr] \Bigl[ \mathds{C}_\text{EH}^{\pm ji}(\omega') e^{\text{i} \omega' \tau} + \mathds{C}_\text{EH}^{\mp ji}(\omega') e^{-\text{i} \omega' \tau} \Bigr]^{*} \Bigr) \Bigg\}
\label{eq:var_u}
\end{align}
with 
\begin{align}
\mathds{C}_\text{AB}^{\pm ij}(\omega) & = \Big \llangle \hat{A}^{+}_i (\mathbf{r}, \omega) \hat{B}^{-}_j (\mathbf{r}', \omega) \Big \rrangle .
\label{eq:CABpmij}
\end{align}
Note that indices $i$ and $j$ indicate the dependence on the coordinates $\mathbf{r}$ and $\mathbf{r}'$, respectively. Because of stationarity the variance only depends on the time difference $\tau = t - t'$. Note that the relation
\begin{align}
\mathds{C}_\text{AB}^{\mp ij}(\omega) & = {\mathds{C}_\text{BA}^{\mp ji}(\omega)}^{*}
\end{align}
was used as well. Now, let me do the same calculation for the intensity fluctuations
\begin{align}
\text{Var}_I & = \Big \llangle I(\mathbf{r}, t) I(\mathbf{r}', t') \Big \rrangle - \Big \llangle I(\mathbf{r}, t) \Big \rrangle \Big \llangle I(\mathbf{r}', t') \Big \rrangle \notag \\
& = \expval{\hat{E}_i^{-}(\mathbf{r}, t) \hat{E}_j^{+}(\mathbf{r}', t')} \expval{\hat{E}_j^{-}(\mathbf{r}', t') \hat{E}_i^{+}(\mathbf{r}, t)} \notag \\
& \quad + \frac{\mu_0^2}{\varepsilon_0^2} \expval{\hat{H}_i^{-}(\mathbf{r}, t) \hat{H}_j^{+}(\mathbf{r}', t')} \expval{\hat{H}_j^{-}(\mathbf{r}', t') \hat{H}_i^{+}(\mathbf{r}, t)} \notag \\
& \quad + \frac{\mu_0}{\varepsilon_0} \Bigl[ \expval{\hat{E}_i^{-}(\mathbf{r}, t) \hat{H}_j^{+}(\mathbf{r}', t')} \expval{\hat{H}_j^{-}(\mathbf{r}', t') \hat{E}_i^{+}(\mathbf{r}, t)} \notag \\
& \quad + \expval{\hat{H}_i^{-}(\mathbf{r}, t) \hat{E}_j^{+}(\mathbf{r}', t')} \expval{\hat{E}_j^{-}(\mathbf{r}', t') \hat{H}_i^{+}(\mathbf{r}, t)} \Bigr] .
\end{align}
Keeping in mind that the fields with identical frequency sign are uncorrelated, due to the evaluation of correlation functions with either only creation or annihilation operators, I obtain
\begin{align}
\text{Var}_I & = \int_0^\infty \frac{\mathrm{d} \omega}{2 \pi} \int_0^\infty \frac{\mathrm{d} \omega'}{2 \pi} \Bigg\{ \mathds{C}_\text{EE}^{\mp ij}(\omega) \mathds{C}_\text{EE}^{\mp ji}(\omega') + \frac{\mu_0^2}{\varepsilon_0^2} \mathds{C}_\text{HH}^{\mp ij}(\omega) \mathds{C}_\text{HH}^{\mp ji}(\omega') \notag \\
& \quad + \frac{\mu_0}{\varepsilon_0} \Bigr[ \mathds{C}_\text{EH}^{\mp ij}(\omega) \mathds{C}_\text{EH}^{\mp ji}(\omega') + \mathds{C}_\text{HE}^{\mp ij}(\omega) \mathds{C}_\text{HE}^{\mp ji}(\omega') \Bigr] \Bigg\} e^{\text{i} (\omega - \omega') \tau} .
\end{align}

\begin{figure}[hbt]
    \center
    \includegraphics[width=0.5\textwidth]{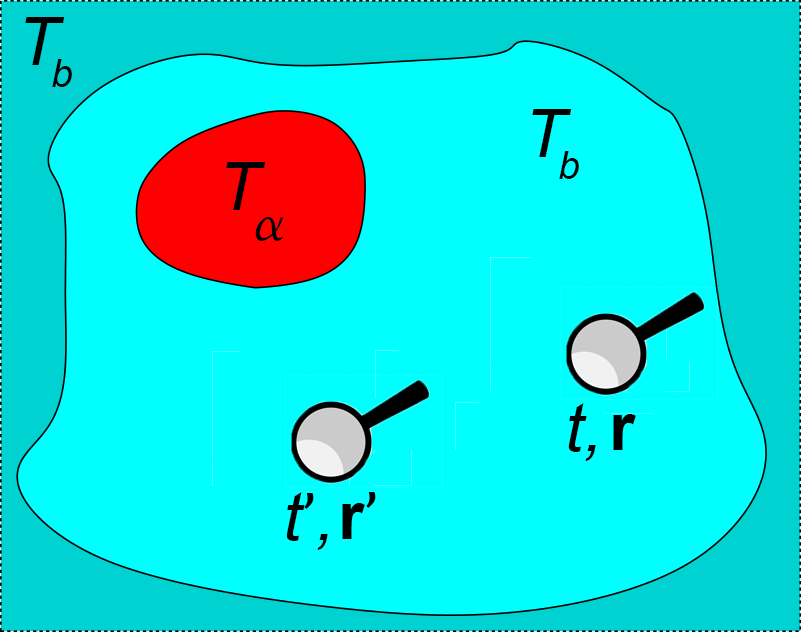}
    \caption{Scheme of the considered system. An arbitrary object with temperature $T_\alpha$ is immersed in an arbitrary background at temperature $T_\text{b}$. Here the environment is sketched as a cavity so that the cavity walls bring the environmental field into local equilibrium with temperature $T_\text{b}$. The calculated variances contain information about the heat radiation of the object and the background measured at $(t, \mathbf{r})$ and $(t', \mathbf{r}')$ (magnifying glasses) independently.}
    \label{fig:scheme}
\end{figure}
Now, I derive the general expressions of $\Big \llangle u \Big \rrangle$, $\Big \llangle I \Big \rrangle$, and their variances for one arbitrary object labeled by $\alpha$ in an arbitrary environment labeled by $\text{b}$ (see Fig.~\ref{fig:scheme}). For this, I use the scattering approach outlined in Ref.~\cite{Herz2} going back to the formalism introduced in Ref.~\cite{Rahi,KruegerEtAlTraceFormulas2012}. The fields and current densities are written in Dirac notation to obtain basis independent formulas. Therein, the current density $\ket{\mathbf{J}}$ is defined by 
\begin{align}
\ket{\mathbf{J}} & = \ket{\mathbf{J}_\text{fl}} + \frac{1}{\text{i} \mu_0 \omega} \mathds{T} \ket{\mathbf{E}_\text{b}}
\label{eq:J_1}
\end{align}
where $\ket{\mathbf{J}_\text{fl}}$ reflects the fluctuational part of the current density and the second contribution shows the induced part. Additionally, the T-operator $\mathds{T}$ is used. The T-operator contains the scattering behavior of heat radiation due to the material. The fields $\ket{\mathbf{F}_{\rm E}} = \ket{\mathbf{E}}$ and $\ket{\mathbf{F}_{\rm H}} = \ket{\mathbf{H}}$ are given by
\begin{align}
\ket{\mathbf{F}_k} & = \ket{\mathbf{F}_{k, \text{b}}} + \text{i} \mu_0 \omega \mathds{G}_{k\text{E}} \ket{\mathbf{J}}
\label{eq:F_1}
\end{align}
with $k \in \{ \text{E,H} \}$ and the Green's function $\mathds{G}$ containing the scattering behavior of heat radiation due to the environment. Then, the correlation function of the fields is
\begin{align}
\Big \llangle \ket{\mathbf{F}_k} \otimes \bra{\mathbf{F}_l} \Big \rrangle & = 2 \hbar \mu_0 \omega^2 \left[ \left[ n_\text{b}(\omega) + 1 \right] \frac{\mathds{G}_{\text{full},kl} - \mathds{G}_{\text{full},lk}^\dagger}{2 \text{i}} + \left[ n_\alpha(\omega) - n_\text{b}(\omega) \right] \mathds{K}_{kl} \right]
\label{eq:FF}
\end{align}
with
\begin{align}
\mathds{G}_{\text{full}, kl} & = \mathds{G}_{kl} + \mathds{G}_{k\text{E}} \mathds{T} \mathds{G}_{\text{E}l}, \\
\mathds{K}_{kl} & = \mathds{G}_{k \text{E}} \boldsymbol{\chi} \mathds{G}^\dagger_{l \text{E}} ,
\label{eq:Kkl}
\end{align}
the Bose-Einstein occupation probability
\begin{align}
n_\gamma(\omega) & = \frac{1}{e^{\frac{\hbar \omega}{k_{\rm B} T_\gamma}} - 1}
\end{align}
with the reduced Planck's constant $\hbar$, the Boltzmann constant $k_{\rm B}$, and the temperature $T_\gamma$ of object $\gamma$ as well as the general susceptibility
\begin{align}
\boldsymbol{\chi} & = \frac{\mathds{T} - \mathds{T}^\dagger}{2 \text{i}} - \mathds{T} \frac{\mathds{G}_\text{EE} - \mathds{G}_\text{EE}^\dagger}{2 \text{i}} \mathds{T}^\dagger .
\end{align}
Keep in mind that $\ket{\mathbf{E}}$ is related to $\mathbf{E}^{+}$ and $\bra{\mathbf{E}}$ is related to $\mathbf{E}^{-}$. That means by interchanging kets and bras, the pre-factor $n_\text{b}(\omega) + 1$ in Eq.~\eqref{eq:FF} simply reduces to $n_\text{b}(\omega)$. The same is true for the magnetic field. With the correlation function in Eq.~\eqref{eq:FF}, the intensity becomes
\begin{align}
\Big \llangle I(\mathbf{r}, t) \Big \rrangle & = \frac{2}{\varepsilon_0} \sum_{k \in \{ \text{E,H}\}} \int_0^\infty \! \frac{\mathrm{d} \omega}{2 \pi} \text{Tr} \Bigr[ \mathds{B}_{kk} (\mathbf{r}, \mathbf{r}, \omega) + \mathds{Q}_{kk} (\mathbf{r}, \mathbf{r}, \omega) \Bigr] .
\label{eq:I}
\end{align}
and its variance is 
\begin{align}
\text{Var}_I & = \frac{4}{\varepsilon_0^2} \sum_{k,l \in \{\text{E,H}\}} \int_0^\infty \! \frac{\mathrm{d} \omega}{2 \pi} \int_0^\infty \! \frac{\mathrm{d} \omega'}{2 \pi} \text{Tr} \Bigl( \Bigr[ \mathds{B}_{kl} (\mathbf{r}, \mathbf{r}', \omega) + \mathds{Q}_{kl} (\mathbf{r}, \mathbf{r}', \omega) \Bigr] \notag \\
& \quad \times \Bigr[ \mathds{B}_{kl}^\dagger (\mathbf{r}, \mathbf{r}', \omega') + \mathds{Q}_{kl}^\dagger (\mathbf{r}, \mathbf{r}', \omega') \Bigr] \Bigr) e^{\text{i} (\omega - \omega') \tau} .
\label{eq:Var_I}
\end{align}
Note, that the energy density always contains vacuum fluctuations whose frequency integrals do not converge, in general, whereas the mean intensity and its variance do not contain vacuum fluctuations. Since I am interested in the evaluation of the thermal contribution of the energy density fluctuations, I will neglect the vacuum contribution in the following. This yields the mean energy density
\begin{align}
\Big \llangle u_\text{th}(\mathbf{r}, t) \Big \rrangle & = 2 \sum_{k \in \{ \text{E,H}\}} \int_0^\infty \! \frac{\mathrm{d} \omega}{2 \pi} \text{Tr} \Bigr[ \text{Re} \left(\mathds{B}_{kk} (\mathbf{r}, \mathbf{r}, \omega) \right) + \mathds{Q}_{kk} (\mathbf{r}, \mathbf{r}, \omega) \Bigr]
\label{eq:u_th}
\end{align}
and its variance
\begin{align}
\text{Var}_{u,\text{th}} & = 8 \sum_{k,l \in \{\text{E,H}\}} \int_0^\infty \! \frac{\mathrm{d} \omega}{2 \pi} \int_0^\infty \! \frac{\mathrm{d} \omega'}{2 \pi} \text{Tr} \Bigl( \text{Re} \left(\mathds{B}_{kl} (\mathbf{r}, \mathbf{r}', \omega) e^{- \text{i} \omega \tau} \right) \notag \\
& \quad + \text{Re} \left(\mathds{Q}_{kl} (\mathbf{r}, \mathbf{r}', \omega) e^{- \text{i} \omega \tau} \right) \Bigr) \Bigl[ \text{Re} \left(\mathds{B}_{lk} (\mathbf{r}', \mathbf{r}, \omega') e^{\text{i} \omega' \tau} \right) \notag \\
& \quad + \text{Re} \left(\mathds{Q}_{lk} (\mathbf{r}', \mathbf{r}, \omega') e^{\text{i} \omega' \tau} \right) \Bigr]
\label{eq:Var_u_th}
\end{align}
using the abbreviations 
\begin{align}
\mathds{B}_{kl} (\mathbf{r}, \mathbf{r}', \omega) & = 2 a_{kl} \hbar k_0^2 n_\text{b} (\omega) \frac{\mathds{G}_{\text{full},kl} (\mathbf{r}, \mathbf{r}', \omega) - \mathds{G}_{\text{full},lk}^\dagger (\mathbf{r}', \mathbf{r}, \omega)}{2 \text{i}} , 
\label{eq:Bkl} \\
\mathds{Q}_{kl} (\mathbf{r}, \mathbf{r}', \omega) & = 2 a_{kl} \hbar k_0^2 \left[ n_\alpha (\omega) - n_\text{b}(\omega) \right] \mathds{K}_{kl} (\mathbf{r}, \mathbf{r}', \omega) ,
\label{eq:Qkl}
\end{align}
and
\begin{align}
a_{kl} & = \begin{cases} 1 & k = l = \text{E} \\ \frac{\mu_0}{\varepsilon_0} & k = l = \text{H} \\ \sqrt{\frac{\mu_0}{\varepsilon_0}} & k \neq l \end{cases} .
\end{align}
Here I introduced the vacuum wave number $k_0 = \omega/c$. Note that Tr$(\mathds{Q} (\mathbf{r}, \mathbf{r}, \omega))$ has only real components.

Let me first consider the special case of the variances at $\mathbf{r}'=\mathbf{r}$ and $\tau=0$ for the pure electric case. If one finds a coordinate system in which $\mathds{M}_\text{EE} (\mathbf{r}) = \int_0^\infty \! \frac{\mathrm{d} \omega}{2 \pi} [\mathds{B}_\text{EE} (\mathbf{r}, \mathbf{r}, \omega) + \mathds{Q}_\text{EE} (\mathbf{r}, \mathbf{r}, \omega)]$ is diagonal, it is possible to decompose this matrix into $\mathds{M}_\text{EE} = \mathds{S} \mathds{D}_\text{EE} \mathds{S}^{-1}$. Here, the diagonal matrix $\mathds{D}_\text{EE}$ contains the eigenvalues $\lambda_\text{EE}$ of $\mathds{M}_\text{EE}$ and the matrix $\mathds{S}$ has the corresponding eigenvectors of $\mathds{M}_\text{EE}$ as its columns. Due to the trace operation, one ends up with
\begin{align}
\Big \llangle I_\text{E} (\mathbf{r}, 0) \Big \rrangle & = \frac{2}{\varepsilon_0} \sum_{i=1}^3 \lambda_{\text{EE},i} (\mathbf{r}) = \frac{\Big \llangle u_\text{th,E}(\mathbf{r}, 0) \Big \rrangle }{\varepsilon_0}
\end{align}
and
\begin{align}
\text{Var}_{u,\text{th,E}} (\mathbf{r}, \mathbf{r}, 0) & = 2 \sum_{i=1}^3 \Big \llangle u_{\text{th,E},i}(\mathbf{r}, 0) \Big \rrangle^2, 
\label{eq:rel1_sum} \\
\text{Var}_{I,\text{E}} (\mathbf{r}, \mathbf{r}, 0) & = \sum_{i=1}^3 \Big \llangle I_{\text{E},i} (\mathbf{r}, 0) \Big \rrangle^2 
\label{eq:rel2_sum} .
\end{align}
When considering isotropic systems like vacuum, all eigenvalues are equal yielding
\begin{align}
\Big \llangle I_\text{E} (\mathbf{r}, 0) \Big \rrangle & = \frac{6}{\varepsilon_0} \lambda_\text{EE} (\mathbf{r}) = \frac{\Big \llangle u_\text{th,E}(\mathbf{r}, 0) \Big \rrangle}{\varepsilon_0}
\end{align}
and
\begin{align}
\text{Var}_{u,\text{th,E}} (\mathbf{r}, \mathbf{r}, 0) & = \frac{2}{3} \Big \llangle u_\text{th,E}(\mathbf{r}, 0) \Big \rrangle^2, 
\label{eq:rel1} \\
\text{Var}_{I,\text{E}} (\mathbf{r}, \mathbf{r}, 0) & = \frac{1}{3} \Big \llangle I_\text{E} (\mathbf{r}, 0) \Big \rrangle^2 .
\label{eq:rel2}
\end{align}
The latter exactly coincides with the result found in \cite{MandelWolf} meaning that intensity fluctuations are on the same order of magnitude as their mean values. Eq.~\eqref{eq:rel1} now shows that the same is true for the thermal energy density fluctuations but with a different pre-factor.

In general, to investigate the second order coherence properties of thermal radiation, it is reasonable to compare the variance of the considered quantity with its mean value. It allows for classifying the non-classical character of light \cite{glauber, aspect} as first measured by the Hanbury Brown-Twiss (HBT) experiment \cite{hbt}. This experiment determined the second order intensity correlation function of a system by using two independent detectors to measure the intensity of a light source for different distances to the source and for time delays between the detectors. The ratio of the second order correlation function and the squared first order correlation function is also called ``complex degree of coherence of second order'' defined by 
\begin{align}
g_{u,\text{th}}^{(2)}(\mathbf{r},\mathbf{r}',\tau) & = \frac{\Big \llangle u_\text{th}(\mathbf{r}, t) u_\text{th}(\mathbf{r}', t') \Big \rrangle}{\Big \llangle u_\text{th}(\mathbf{r}, t) \Big \rrangle \Big \llangle u_\text{th}(\mathbf{r}', t') \Big \rrangle} \notag \\
& = 1 + \frac{\text{Var}_{u,\text{th}} (\mathbf{r}, \mathbf{r}', \tau)}{\Big \llangle u_\text{th}(\mathbf{r}, t) \Big \rrangle \Big \llangle u_\text{th}(\mathbf{r}', t') \Big \rrangle}
\label{eq:g2u}
\end{align}
for the energy density and
\begin{align}
g_I^{(2)}(\mathbf{r},\mathbf{r}',\tau) & = 1 + \frac{\text{Var}_{I} (\mathbf{r}, \mathbf{r}', \tau)}{\Big \llangle I(\mathbf{r}, t) \Big \rrangle \Big \llangle I(\mathbf{r}', t') \Big \rrangle}
\label{eq:g2I}
\end{align}
for the intensity. In principle, a setup like the HBT experiment should also be able to measure the second order intensity correlation function of the system considered in this work. Thermal radiation showing bunching belongs to the class of quasi-classical light. For isotropic objects and environments, one can directly read off, due to Eqs.~\eqref{eq:rel1}-\eqref{eq:rel2}, that 
\begin{align}
g_{u_\text{E},\text{th}}^{(2)}(\mathbf{r},\mathbf{r},0) & = \frac{5}{3} = g_{I_\text{E}}^{(2)}(\mathbf{r},\mathbf{r},0) + \frac{1}{3}
\end{align}
when only considering electric contributions. This also defines the maximum value of $g_\text{E}^{(2)}$ for both, spatial and temporal bunching. 

\section{Numerical Results}

At first, let me validate the main results of Eq.~\eqref{eq:I}-\eqref{eq:Var_u_th} and Eq.~\eqref{eq:g2u}-\eqref{eq:g2I} by retrieving the well known result for the case of pure vacuum \cite{wolf}. Additionally, I want to apply it to two more complex examples of practical interest for experiments, namely a half-space and a sphere, that have not been investigated with respect to the variances of energy density and intensity, yet. In the following, I will use two different types of materials: SiC and gold. SiC as a dielectric material can be modeled by a Lorentz-Oscillator \cite{Palik}
\begin{align}
\varepsilon_\text{SiC} (\omega) & = \varepsilon_\infty \frac{\omega_l^2 - \omega^2 - \text{i} \Gamma \omega}{\omega_t^2 - \omega^2 - \text{i} \Gamma \omega}
\end{align}
with $\varepsilon_\infty = 6.7$, $\omega_l = 1.827 \times 10^{14}$ rad/s, $\omega_t = 1.495 \times 10^{14}$ rad/s, and $\Gamma = 0.9 \times 10^{12}$ rad/s. For gold I employ the Drude model \cite{Ordal}
\begin{align}
\varepsilon_\text{Au} (\omega) & = \varepsilon_\infty - \frac{\omega_p^2}{\omega^2 + \text{i} \Gamma \omega}
\end{align}
with $\varepsilon_\infty = 8.344$, $\omega_p = 1.372 \times 10^{16}$ rad/s, and $\Gamma = 4.059 \times 10^{13}$ rad/s.

\subsection{Energy density and intensity fluctuations of vacuum} \label{ch:ed_vac}

In the simplest case of considering black-body radiation in vacuum, one can set $\mathds{T}=0$ and, thus, obtain $\mathds{Q}_{kl}=0$ because no object is involved. Then, if one is only interested in the temporal coherence, the spatial arguments become identical. The Green's functions become $\mathds{G}_{\text{full}, kl} (\mathbf{r}, \mathbf{r}) = \mathds{G}_{kl} (\mathbf{r}, \mathbf{r})$ yielding for Eq.~\eqref{eq:Bkl}
\begin{align}
\mathds{B}_{kl} (\mathbf{r}, \mathbf{r}, \omega) & = 2 a_{kl} \hbar k_0^2 n_\text{b}(\omega) \times \begin{cases} \frac{\omega}{6 \pi c} \mathds{1} & k = l \\ 0 & k \neq l \end{cases} .
\end{align}
By using this in Eqs.~\eqref{eq:I}-\eqref{eq:Var_u_th}, it is straightforward to derive the desired results for the mean values and variances 
\begin{align}
\Big \llangle u(\mathbf{r}, t) \Big \rrangle & = \varepsilon_0 \Big \llangle I(\mathbf{r}, t) \Big \rrangle = \frac{\pi^2 k_\text{B}^4 T_\text{b}^4}{15 \hbar^3 c^3}
\label{eq:u_I_vac}
\end{align}
and
\begin{align}
\text{Var}_{u,\text{th}} (\mathbf{r}, \mathbf{r}, \tau) & = \frac{12 \hbar^2}{\pi^4 c^6 \tau_\text{b}^8} \biggl[\text{Re} \left( \zeta \left(4, 1 - \text{i} \frac{\tau}{\tau_\text{b}} \right) \right) \biggr]^2 , 
\label{eq:Du_vac} \\
\text{Var}_{I} (\mathbf{r}, \mathbf{r}, \tau) & = \frac{6 \hbar^2}{\varepsilon_0^2 \pi^4 c^6 \tau_\text{b}^8} \bigg| \zeta \left(4, 1 - \text{i} \frac{\tau}{\tau_\text{b}} \right) \bigg|^2
\label{eq:DI_vac}
\end{align}
which coincide with the corresponding expression in \cite{wolf}. Here, 
\begin{align}
\zeta(x,y) & = \sum_{n=0}^\infty \frac{1}{(n+y)^x}
\end{align}
is the Hurwitz zeta function, $T_\text{b}$ is the vacuum background temperature, and I defined the vacuum's coherence time
\begin{align}
\tau_\text{b} & = \frac{\hbar}{k_\text{B} T_\text{b}}.
\end{align}
Comparing the full variances with the mean values, I find
\begin{align}
\text{Var}_{u,\text{th}} (\mathbf{r}, \mathbf{r}, 0) & = \frac{1}{3} \Big \llangle u(\mathbf{r}, t) \Big \rrangle^2, 
\label{eq:u_vac} \\
\text{Var}_{I} (\mathbf{r}, \mathbf{r}, 0) & = \frac{1}{6} \Big \llangle I(\mathbf{r}, t) \Big \rrangle^2.
\label{eq:I_vac}
\end{align}
However, the pure electric or magnetic contributions fulfill the relations of Eqs.~\eqref{eq:rel1}-\eqref{eq:rel2} because both contribute equally to Eqs.~\eqref{eq:u_vac}-\eqref{eq:I_vac}. The factor of $2$ is missing in Eqs.~\eqref{eq:u_vac}-\eqref{eq:I_vac} compared to Eqs.~\eqref{eq:rel1}-\eqref{eq:rel2} when only considering one contribution, e.g.\ only the electric one. Thereby, I retrieve the results of Ref.~\cite{MandelWolf} for vacuum. 

\begin{figure}[hbt]
    \center
    \includegraphics[width=0.7\textwidth]{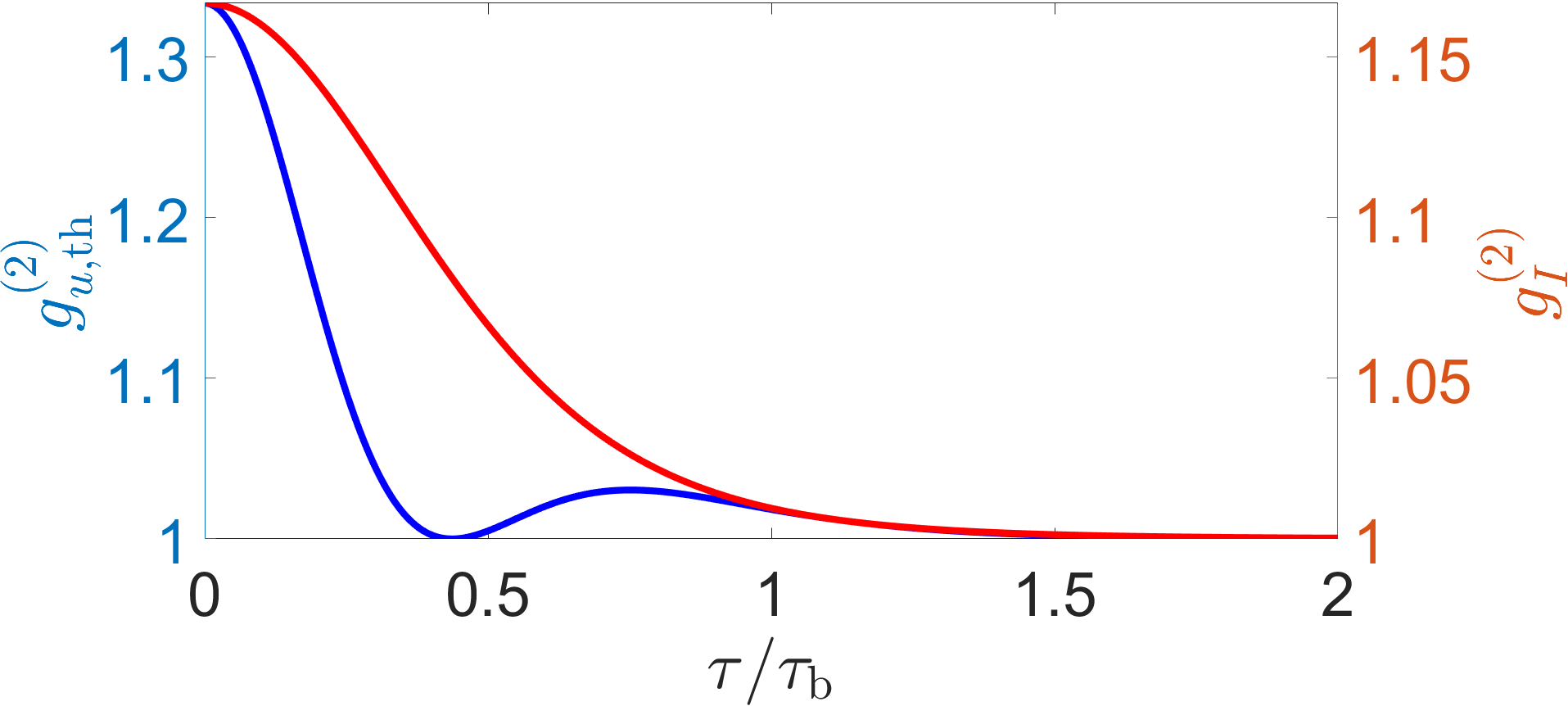}
    \caption{$g^{(2)}$ function of the thermal contribution of the energy density (blue) and of the intensity (red) with respect to the normalized time delay $\tau/\tau_\text{b}$ for $\tau_\text{b} = 25.5$ fs at $T_\text{b}=300$ K.}
    \label{fig:vac}
\end{figure}

The $g^{(2)}$ function for the energy density and the intensity are depicted in Fig.~\ref{fig:vac}. For both quantities one can clearly see the bunching character of thermal vacuum radiation and the loss of coherence at a time delay of $\tau_\text{b} = 25.5$ fs at $T_\text{b}=300$ K.

\subsection{Energy density and intensity fluctuations above a planar substrate}

A more sophisticated problem is the calculation of the energy density and intensity fluctuations at distance $d$ above a half-space. The half-space is assumed to cover the whole x-y space and the space $z<0$. I only take non-magnetic, homogeneous, and isotropic materials into account for the half-space. The temperature of the half-space is $T_\alpha$ and the one of the background $T_\text{b}$. The Green's function for the whole system is well known \cite{Sipe} and can be separated into a vacuum part and a scattered contribution due to reflections at the surface of the half-space within the planar wave base as
\begin{align}
\mathds{G}_\text{full,EE} (\mathbf{r}, \mathbf{r}') & = \int \frac{\text{d}^2 k_\perp}{(2 \pi)^2} e^{\text{i} \mathbf{k}_\perp \mathbf{x}_\perp} \left[ \mathds{G}_\text{vac,EE} (\mathbf{k}_\perp, z, z') + \mathds{G}_\text{scat,EE} (\mathbf{k}_\perp, z, z') \right] .
\label{eq:Gf_plane}
\end{align}
Note that I assume $\mathbf{r}$ and $\mathbf{r}'$ being outside of the half-space. The remaining Green's functions $\mathds{G}_\text{EH/HE/HH}$ can be obtained by exploiting the duality relations between the desired Green's function and $\mathds{G}_\text{EE}$. The explicit expressions can be found in appendix \ref{appA}. Then, it is easy to find the $\mathds{B}$ matrices using Eq.~\eqref{eq:Bkl}
\begin{align}
\mathds{B}_{\text{EE/HH}} (\mathbf{r}, \mathbf{r}, \omega) & = \hbar k_0^3 n_\text{b}(\omega) \sum_{j \in \{\perp,\parallel\}} \left(\frac{1}{6 \pi} (1 + \delta_{j \perp}) +  I_{\text{p/s},j} \right) \mathds{A}_j , 
\label{eq:B1_hs} \\
\mathds{B}_{\text{EH/HE}} (\mathbf{r}, \mathbf{r}, \omega) & = \text{i} \hbar k_0^3 n_\text{b}(\omega) I_\text{mix} \mathds{A}_\text{mix} , 
\end{align}
where I introduced the matrices 
\begin{align}
\mathds{A}_\perp & = \frac{1}{2} \left(\mathbf{e}_x \otimes \mathbf{e}_x + \mathbf{e}_y \otimes \mathbf{e}_y \right) , \\
\mathds{A}_\parallel & = \mathbf{e}_z \otimes \mathbf{e}_z, 
\end{align}
and
\begin{align}
\mathds{A}_\text{mix} & = \mathbf{e}_x \otimes \mathbf{e}_y - \mathbf{e}_y \otimes \mathbf{e}_x
\end{align}
as well as $I_{\text{p/s},j}$ and $I_\text{mix}$ defined in Eqs.~\eqref{eq:I_ps_perp}-\eqref{eq:I_mix}. Note that $\mathds{B}_\text{HE} = \mathds{B}_\text{EH}^\dagger$. The first term in \eqref{eq:B1_hs} corresponds to the vacuum contribution, whereas the second part describes the reflected parts. The mixed Green's functions possess no vacuum part as in the case of pure vacuum. 

To obtain the remaining $\mathds{Q}$ matrices defined in Eq.~\eqref{eq:Qkl}, I expand the Green's function and T-operators in $\mathds{K}$ as defined in Eq.~\eqref{eq:Kkl} in the plane wave basis as well. These expressions are also well known in literature \cite{Sipe, Herz3} so that I get with Eq.~\eqref{eq:Qkl}
\begin{align}
\mathds{Q}_\text{EE/HH} (\mathbf{r}, \mathbf{r}, \omega) & = \hbar k_0^3 \left[ n_\alpha(\omega) - n_\text{b}(\omega) \right] \sum_{j \in \{\perp, \parallel\}} K_{\text{p/s},j} \mathds{A}_j , \\
\mathds{Q}_\text{EH/HE} (\mathbf{r}, \mathbf{r}, \omega) & = \mp \hbar k_0^3 \left[ n_\alpha(\omega) - n_\text{b}(\omega) \right] \left[ K^\text{pr}_\text{mix} \pm \text{i} K^\text{ev}_\text{mix} \right] \mathds{A}_\text{mix}
\end{align}
with $K_{\text{p/s},j}$, $K_\text{mix}^\text{pr}$, and $K_\text{mix}^\text{ev}$ defined in Eqs.~\eqref{eq:K_ps_perp}-\eqref{eq:K_ev_mix}. Inserting these in Eqs.~\eqref{eq:I}-\eqref{eq:Var_u_th}, yields the mean values
\begin{align}
\Big \llangle u_\text{th}(\mathbf{r}, t) \Big \rrangle & = \varepsilon_0 \sum_{k \in \{\text{E,H}\}} \sum_{j \in \{\perp,\parallel\}} \left[ \Gamma^\text{eq}_{k,j} (0) + \Gamma^\text{leq}_{k,j} (0) \right] = \varepsilon_0 \Big \llangle I(\mathbf{r}, t) \Big \rrangle 
\label{eq:u_hs}
\end{align}
and the corresponding variances
\begin{align}
\text{Var}_{u,\text{th}} (\mathbf{r}, \mathbf{r}, \tau) & = \varepsilon_0^2 \sum_{k \in \{\text{E,H}\}} \sum_{j \in \{\perp,\parallel\}} (1 + \delta_{j \parallel}) \text{Re}\left( \Gamma^\text{eq}_{k,j} (\tau) + \Gamma^\text{leq}_{k,j} (\tau) \right)^2 \notag \\
& \quad + 2 \varepsilon_0^2 \text{Im}\Bigl(\Gamma^\text{eq}_\text{mix} (\tau) - \Gamma^\text{leq}_\text{mix,ev} (\tau) \Bigr)^2 + 2 \varepsilon_0^2 \text{Re}\Bigl(\Gamma^\text{leq}_\text{mix,pr} (\tau) \Bigr)^2 , 
\label{eq:var_u_hs} \\
\text{Var}_{I} (\mathbf{r}, \mathbf{r}, \tau) & = \frac{1}{2} \sum_{k \in \{\text{E,H}\}} \sum_{j \in \{\perp,\parallel\}} (1 + \delta_{j \parallel}) \Big| \Gamma^\text{eq}_{k,j} (\tau) + \Gamma^\text{leq}_{k,j} (\tau) \Big|^2 \notag \\
& \quad + \Big| \Gamma^\text{eq}_\text{mix} (\tau) - \Gamma^\text{leq}_\text{mix,ev} (\tau) \Big|^2 + \Big| \Gamma^\text{leq}_\text{mix,pr} (\tau) \Big|^2 .
\label{eq:var_I_hs}
\end{align}
The $\Gamma$ integrals are defined in Eqs.~\eqref{eq:G_leq_EH}-\eqref{eq:G_eq_mix}. These variances fulfill the relations in Eqs.~\eqref{eq:rel1_sum}-\eqref{eq:rel2_sum} when taking into account that there are two directions parallel to the half-space's surface which equally contribute to the energy density and the intensity, then
\begin{align}
\text{Var}_{u,\text{th,E}} (\mathbf{r}, \mathbf{r}, 0) & = 2 \sum_{j \in \{\text{x,y,z}\}} \Big \llangle u_{\text{th,E},j}(\mathbf{r}, t) \Big \rrangle^2 \notag \\
& = 2 \left( 2 \Big \llangle u_{\text{th,E},x}(\mathbf{r}, t) \Big \rrangle^2 + \Big \llangle u_{\text{th,E},z}(\mathbf{r}, t) \Big \rrangle^2 \right), \\
\text{Var}_{I,\text{E}} (\mathbf{r}, \mathbf{r}, 0) & = \sum_{j \in \{\text{x,y,z}\}} \Big \llangle I_{\text{E},j}(\mathbf{r}, t) \Big \rrangle^2 = 2 \Big \llangle I_{\text{E},x}(\mathbf{r}, t) \Big \rrangle^2 + \Big \llangle I_{\text{E},z}(\mathbf{r}, t) \Big \rrangle^2 
\end{align}
with 
\begin{align}
\Big \llangle u_\text{th,E,x/y/z}(\mathbf{r}, t) \Big \rrangle & = \varepsilon_0 \left[ \Gamma^\text{eq}_\text{E,x/y/z} (0) + \Gamma^\text{leq}_\text{E,x/y/z} (0) \right] = \varepsilon_0 \Big \llangle I_\text{E,x/y/z}(\mathbf{r}, t) \Big \rrangle
\end{align}
and $\Gamma^\text{(l)eq}_\text{E,x} = \Gamma^\text{(l)eq}_\text{E,y} = \Gamma^\text{(l)eq}_{\text{E},\perp} / 2$.

To validate the results in Eqs.~\eqref{eq:u_hs}-\eqref{eq:var_I_hs}, let me investigate their asymptotic behavior for some limiting cases. If the position $\mathbf{r}$ of the observer is far removed from the half space ($d \rightarrow \infty$), one can safely neglect all evanescent parts of the $I$ integrals. Additionally, the scattering terms of the propagating contributions containing the complex exponential functions $e^{2 \text{i} k_z d}$ tend to show decaying features due to the rapid oscillations. The resulting integrals are summarized in Eqs.~\eqref{eq:K_ps_perp}-\eqref{eq:G_mix}. In this limit I obtain the mean values
\begin{align}
\Big \llangle u_\text{th}(\mathbf{r}, t) \Big \rrangle & = \frac{\pi^2 k_\text{B}^4}{30 c^3 \hbar^3} \left[ T_\alpha^4 + T_\text{b}^4 \right] - \frac{\hbar}{4 \pi^2} \int_0^\infty \! \mathrm{d} \omega k_0^2 [n_\alpha (\omega) - n_\text{b} (\omega)] \int_0^{k_0} \text{d} k_\perp \frac{k_\perp}{k_z} \left( |r_\text{s}(\mathbf{k}_\perp)|^2 + |r_\text{p}(\mathbf{k}_\perp)|^2 \right) = \epsilon_0 \Big \llangle I(\mathbf{r}, t) \Big \rrangle 
\label{eq:u_ult}
\end{align}
and the variances
\begin{align}
\text{Var}_{u,\text{th}} (\mathbf{r}, \mathbf{r}, \tau) & = \varepsilon_0^2 \bigg\{ \left[ \text{Re} \left(\Gamma_{\text{vac},1} (\tau) - \Gamma^{d \rightarrow \infty}_{\text{E},\perp} (\tau) \right) \right]^2 + 2 \left[ \text{Re} \left( \frac{1}{2} \Gamma_{\text{vac},1} (\tau) - \Gamma^{d \rightarrow \infty}_{\text{E},z} (\tau) \right) \right]^2 + \left[ \text{Re} \left( \Gamma_{\text{vac},1} (\tau) - \Gamma^{d \rightarrow \infty}_{\text{H},\perp} (\tau) \right) \right]^2 \notag \\
& \quad + 2 \left[ \text{Re} \left( \frac{1}{2} \Gamma_{\text{vac},1} (\tau) - \Gamma^{d \rightarrow \infty}_{\text{H},z} (\tau) \right) \right]^2 + 2 \left[ \text{Re} \left( \Gamma_{\text{vac},2} (\tau) - \Gamma^{d \rightarrow \infty}_\text{mix,pr} (\tau) \right) \right]^2 \Bigg\}
\label{eq:var_u_ult} \\
\text{Var}_{I} (\mathbf{r}, \mathbf{r}, \tau) & = \frac{1}{2} \Big| \Gamma_{\text{vac},1} (\tau) - \Gamma^{d \rightarrow \infty}_{\text{E},\perp} (\tau) \Big|^2 + \bigg| \frac{1}{2} \Gamma_{\text{vac},1} (\tau) - \Gamma^{d \rightarrow \infty}_{\text{E},z} (\tau) \bigg|^2 + \frac{1}{2} \Big| \Gamma_{\text{vac},1} (\tau) - \Gamma^{d \rightarrow \infty}_{\text{H},\perp} (\tau) \Big|^2 \notag \\
& \quad + \bigg| \frac{1}{2} \Gamma_{\text{vac},1} (\tau) - \Gamma^{d \rightarrow \infty}_{\text{H},z} (\tau) \bigg|^2 + \Big| \Gamma_{\text{vac},2} (\tau) - \Gamma^{d \rightarrow \infty}_\text{mix,pr} (\tau) \Big|^2 .
\label{eq:var_I_ult}
\end{align}
See Eqs.~\eqref{eq:G_vac1}-\eqref{eq:G_inf_mix} for the abbreviations used in the variance expressions. This already displays that in thermal equilibrium $T_\alpha = T_\beta$ I will obtain the vacuum results because $\Gamma_{\text{vac},1}$ will be the only non-vanishing contribution. In the case of a black-body, which means to set $|r_\text{s/p}(\mathbf{k}_\perp)|^2=0$, additionally $\Gamma_{\text{vac},2}$ will contribute to the variances yielding
\begin{align}
\Big \llangle u_\text{th}(\mathbf{r}, t) \Big \rrangle & = \frac{\pi^2 k_\text{B}^4}{30 c^3 \hbar^3} \left[ T_\alpha^4 + T_\text{b}^4 \right] = \epsilon_0 \Big \llangle I(\mathbf{r}, t) \Big \rrangle 
\label{eq:u_hs_bb}
\end{align}
and
\begin{align}
\text{Var}_{u,\text{th}} (\mathbf{r}, \mathbf{r}, \tau) & = \varepsilon_0^2 \Big\{ 3 \left[ \text{Re} \left(\Gamma_{\text{vac},1} (\tau) \right) \right]^2 + 2 \left[ \text{Re} \left( \Gamma_{\text{vac},2} (\tau) \right) \right]^2 \Big\} \\
\text{Var}_{I} (\mathbf{r}, \mathbf{r}, \tau) & = \frac{3}{2} \Big| \Gamma_{\text{vac},1} (\tau) \Big|^2 + \Big| \Gamma_{\text{vac},2} (\tau) \Big|^2 .
\end{align}
Especially Eq.~\eqref{eq:u_hs_bb} shows that both, the black-body half space and the vacuum background, contribute evenly to the energy density. Due to the additional term in both variances, the $g^{(2)}$ functions will be larger compared to the pure vacuum case if $T_\alpha \neq T_\beta$ holds. In the opposite case of a perfect conductor, which means setting $|r_\text{s/p}(\mathbf{k}_\perp)|^2=1$, I obtain 
\begin{align}
\Big \llangle u_\text{th}(\mathbf{r}, t) \Big \rrangle & = \frac{\pi^2 k_\text{B}^4 T_\text{b}^4}{15 c^3 \hbar^3} = \epsilon_0 \Big \llangle I(\mathbf{r}, t) \Big \rrangle 
\end{align}
and the variances
\begin{align}
\text{Var}_{u,\text{th}} (\mathbf{r}, \mathbf{r}, \tau) & = 12 \left[ \text{Re} \left(\frac{k_\text{B}^4 T_\text{b}^4}{\pi^2 c^3 \hbar^3} \zeta \left(4, 1 - \text{i} \frac{\tau}{\tau_\text{b}} \right) \right) \right]^2 + 2 \left[ \text{Re} \left( \frac{3 k_\text{B}^4}{4 \pi^2 c^3 \hbar^3} \left[ T_\alpha^4 \zeta \left(4, 1 - \text{i} \frac{\tau}{\tau_\alpha} \right) - T_\text{b}^4 \zeta \left(4, 1 - \text{i} \frac{\tau}{\tau_\text{b}} \right) \right] \right) \right]^2 \\
\text{Var}_{I} (\mathbf{r}, \mathbf{r}, \tau) & = 6 \bigg| \frac{k_\text{B}^4 T_\text{b}^4}{\varepsilon_0 \pi^2 c^3 \hbar^3} \zeta \left(4, 1 - \text{i} \frac{\tau}{\tau_\text{b}} \right) \bigg|^2 + \bigg| \frac{3 k_\text{B}^4}{4 \pi^2 \varepsilon_0 c^3 \hbar^3} \left[ T_\alpha^4 \zeta \left(4, 1 - \text{i} \frac{\tau}{\tau_\alpha} \right) - T_\text{b}^4 \zeta \left(4, 1 - \text{i} \frac{\tau}{\tau_\text{b}} \right) \right] \bigg|^2 .
\end{align}
The mean values become the vacuum results, again. This is due to Kirchhoff's law stating that a perfect conductor who reflects all incoming radiation would also not emit any radiation. The variances also show an additional contribution which reflects the presence of the conducting half space. Again, the $g^{(2)}$ functions will be larger compared to the pure vacuum case if $T_\alpha \neq T_\beta$ holds.

Let me conclude by considering the quasi-static limit $c \rightarrow \infty$ to investigate the case being in the vicinity of the half space. This corresponds to setting $r_\text{s}(\mathbf{k}_\perp)=0$ and $r_\text{p}(\mathbf{k}_\perp)=(\varepsilon - 1)/(\varepsilon + 1)$. The resulting integrals are listed in Eqs.~\eqref{eq:K_qs_p}-\eqref{eq:G_qs_mix}. Note that I only take into account the leading expressions with respect to $1/d$. This ends up in the mean values
\begin{align}
\Big \llangle u_\text{th}(\mathbf{r}, t) \Big \rrangle & = \frac{\hbar}{8 \pi^2 d^3} \int_0^\infty \! \mathrm{d} \omega \left[ 2 n_\alpha (\omega) - n_\text{b} (\omega) \right] \frac{\text{Im}(\varepsilon)}{|\varepsilon + 1|^2} = \varepsilon_0 \Big \llangle I(\mathbf{r}, t) \Big \rrangle 
\end{align}
and variances
\begin{align}
\text{Var}_{u,\text{th}} (\mathbf{r}, \mathbf{r}, \tau) & = \frac{3}{4} \left( \frac{\hbar}{8 \pi^2 d^3} \int_0^\infty \! \mathrm{d} \omega \left[ 2 n_\alpha (\omega) - n_\text{b} (\omega) \right] \frac{\text{Im}(\varepsilon)}{|\varepsilon + 1|^2} \cos(\omega \tau) \right)^2 , \\
\text{Var}_{I} (\mathbf{r}, \mathbf{r}, \tau) & = \frac{3}{8} \bigg| \frac{\hbar}{8 \varepsilon_0 \pi^2 d^3} \int_0^\infty \! \mathrm{d} \omega \left[ 2 n_\alpha (\omega) - n_\text{b} (\omega) \right] \frac{\text{Im}(\varepsilon)}{|\varepsilon + 1|^2} e^{\text{i} \omega \tau} \bigg|^2 .
\end{align}
Note that I neglected the vacuum contribution in this case because it becomes negligibly small compared to the quasi-static limit approach of the scattered part and the half space contribution. Also note that the well-known $1/d^3$ proportionality emerges. In the equilibrium case of $T_\alpha = T_\text{b}$ only the reflection at the half space surface is contributing.

\begin{figure}[hbt]
    \center
    \includegraphics[width=0.7\textwidth]{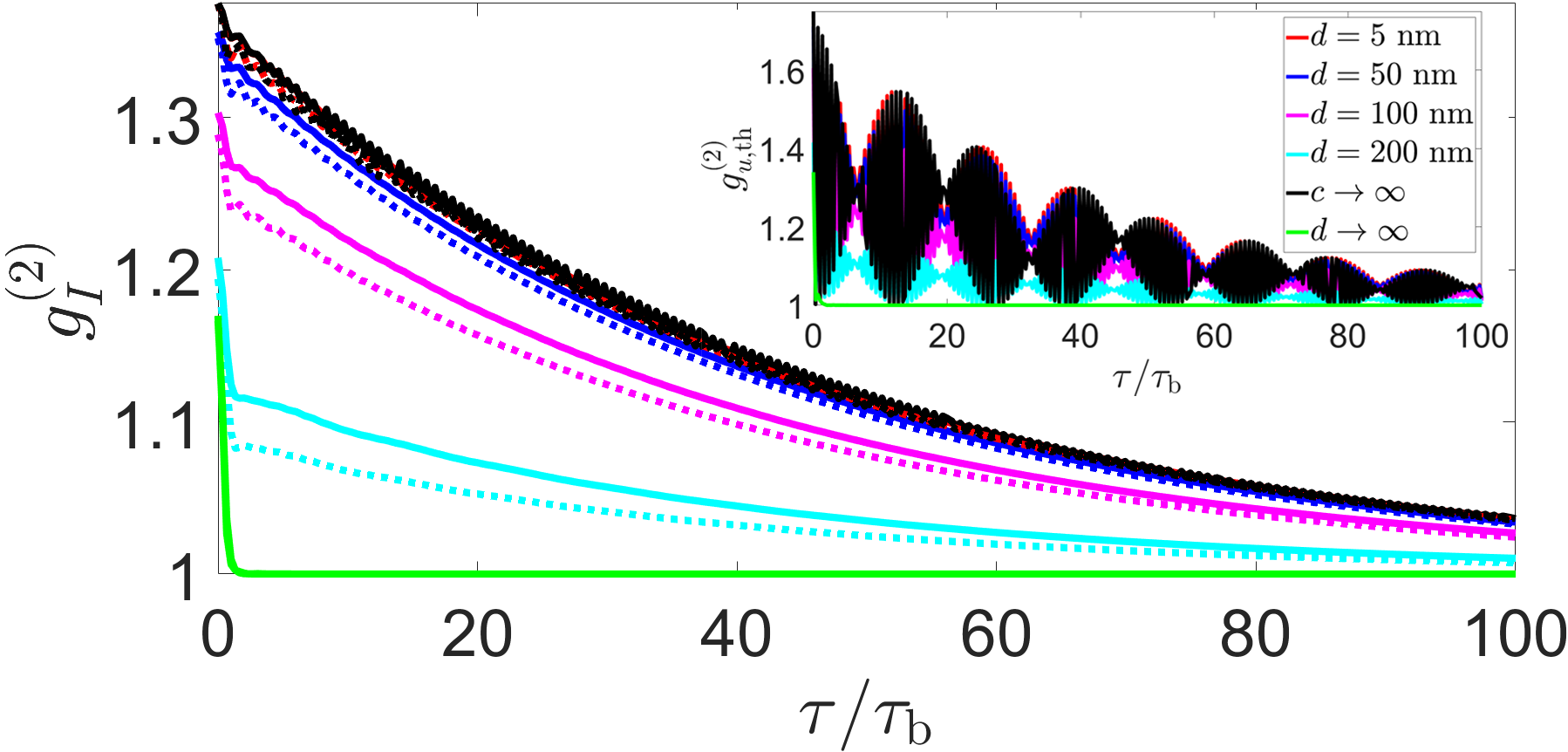}
    \caption{$g_I^{(2)}$ function of the intensity for different distances $d$ to the substrate's surface and for the two limits $c,d \rightarrow \infty$ in global equilibrium conditions ($T_\alpha = T_\text{b}$, dashed lines) as well as in local equilibrium conditions ($T_\alpha \neq T_\text{b}$, solid lines) with respect to the normalized time delay $\tau/\tau_\text{b}$. Inset: $g_{u,\text{th}}^{(2)}$ function of the thermal contribution of the energy density in local equilibrium for the same distances.}
    \label{fig:SiC}
\end{figure}

Let me now move on to some numerical evaluations of the half space example. The resulting $g^{(2)}$ functions are plotted in Fig.~\ref{fig:SiC} for SiC employing the temperatures $T_\alpha = 350$ K and $T_\text{b} = 300$ K.
As one would expect, the coherence time $\tau_c$, which replaces $\tau_\text{b}$ for this geometry, now depends on the distance $d$ between substrate and observation point and exceeds the vacuum value by at least two orders of magnitude, even for distances like $d=200$ nm in agreement with Ref.~\cite{rhe_flucs}. The larger this distance, the shorter becomes the coherence time $\tau_c$ and gets close to $\tau_\text{b}$ again in the far-field regime as shown in Eqs.~\eqref{eq:u_ult}-\eqref{eq:var_I_ult} or coincides with it in the equilibrium case of $T_\alpha=T_\text{b}$. The large values $\tau_c \gg \tau_\text{b}$ in the near-field regime exist because SiC possesses two resonance frequencies that become very pronounced in the near field spectrum. These are the SPhP resonance frequency and the transverse optical phonon (TOP) resonance frequency. The SPhP mode is more pronounced in the electric contribution compared to the TOP mode that dominates the spectrum of the magnetic contribution \cite{Dong,Herz5}. Due to the quasi-monochromatic distribution of the spectrum around these two frequencies the correlation time increases. For larger distances $d$ the effect of these evanescent waves decreases and the amplitude of the $g^{(2)}$ functions decreases as well. This is most pronounced for $d>50$ nm since below this distance the curves almost overlap. There is also an increasing drop of amplitude at very short time delays for growing distances $d$ for the same reason. This drop happens for time delays on the order of $\tau_\text{b}$. Therefore, this drop can be connected to the degrading of the quasi-monochromatic spectrum into a black-body one like in vacuum for growing distances shown here by the $d \rightarrow \infty$ approximation taken from Eqs.~\eqref{eq:u_ult}-\eqref{eq:var_I_ult}. Thus, this $g^{(2)}$ function seems to be an overlap of the one corresponding to the vacuum case that I discussed previously, which explains the drop emerging for larger distances, and the one for the quasi-monochromatic spectrum dominated by the SPhP mode, which explains the large $\tau_c$ values for smaller distances. In the case of $d=5$ nm one can observe a good agreement with the quasi-static limit approximation which basically neglects any vacuum contribution so that one is left with the influence of the quasi-monochromatic spectrum instead of an overlap of it with the black-body case. For the chosen distances $d$, the values for the global equilibrium situation ($T_\alpha = T_\text{b}$) are always smaller than for the local equilibrium case ($T_\alpha \neq T_\text{b}$) due to the missing $\mathds{K}$ matrix contribution. Interestingly, the initial values of the global and local equilibrium conditions for $\tau = 0$ approach each other for increasing distances $d$. Compared to $g_{I}^{(2)}$, $g_{u,\text{th}}^{(2)}$ behaves qualitatively identical regarding the above mentioned points but with a wave-like character. $g_{I}^{(2)}$, however, behaves like the average of $g^{(2)}_{u,\text{th}}$ due to the absolute value. 

\subsection{Energy density and intensity fluctuations in the vicinity of a sphere}

Finally, let me apply the here developed theory to a single sphere of radius $R$ immersed in vacuum for which I compute the mean values and variances of the energy density and intensity at distance $r$ to the center of the sphere. Again, the fully electric Green's function $\mathds{G}_\text{full,EE}$ can be decomposed into a vacuum part and a scattering part due to reflections at the sphere's surface like in Eq.~\eqref{eq:Gf_plane} but for a different geometrical basis. For that I use the notation \cite{KruegerEtAlTraceFormulas2012}
\begin{align}
\mathds{G}_\text{vac,EE} (\mathbf{r}, \mathbf{r}') & = \text{i} \sum_{P,l,m} \mathbf{E}_{P,l,m}^\text{out} (\mathbf{r}) \otimes \mathbf{E}_{P,l,-m}^\text{reg} (\mathbf{r}') , \\
\mathds{G}_\text{scat,EE} (\mathbf{r}, \mathbf{r}') & = \text{i} \sum_{P,l,m} \mathcal{T}_l^P \mathbf{E}_{P,l,m}^\text{out} (\mathbf{r}) \otimes \mathbf{E}_{P,l,-m}^\text{reg} (\mathbf{r}') .
\end{align}
Here, $P \in \{\text{M,N}\}$ corresponds to the two different wave vector solutions to the general Helmholtz equation applied to spherical waves, $l \geq 1$ denotes the multipole order, and $-l \leq m \leq l$ characterizes the multipole index. Note that the scattering at the sphere's surface manifests in $\sigma(m)=-m$ in the second vector of $\mathds{G}_\text{scat,EE}$. $\mathcal{T}$ denotes the T-operator applied to the spherical basis being diagonal for all indices and independent of $m$. Please, find the expressions for the vector functions $\mathbf{E}$ and the T-operator $\mathcal{T}$ in appendix \ref{appB}. Inserting the Green's functions in Eq.~\eqref{eq:Bkl} and \eqref{eq:Qkl} yields the $\mathds{B}$ matrices
\begin{align}
\mathds{B}_{\text{EE/HH}} (\mathbf{r}, \mathbf{r}, \omega) & = \hbar k_0^3 n_\text{b} (\omega) \left[ \frac{1}{6 \pi} \mathds{1} + \sum_{l=1}^\infty \frac{2 l + 1}{4 \pi} \text{Re} \left( \mathcal{B}^\text{E/H}_{l,\vartheta \varphi} \left[ \mathbf{e}_\vartheta \otimes \mathbf{e}_\vartheta + \mathbf{e}_\varphi \otimes \mathbf{e}_\varphi \right] + \mathcal{B}^\text{E/H}_{l,r}  \mathbf{e}_r \otimes \mathbf{e}_r \right) \right], 
\label{eq:B_EE_HH_sp} \\
\mathds{B}_{\text{EH}} (\mathbf{r}, \mathbf{r}, \omega) & = - \text{i} \hbar k_0^3 n_\text{b} (\omega) \sum_{l=1}^\infty \frac{2 l + 1}{8 \pi} \mathcal{B}^\text{mix}_{l,\vartheta \varphi} \left[ \mathbf{e}_\vartheta \otimes \mathbf{e}_\varphi + \mathbf{e}_\varphi \otimes \mathbf{e}_\vartheta \right] ,
\end{align}
and the $\mathds{Q}$ matrices
\begin{align}
\mathds{Q}_\text{EE/HH} (\mathbf{r}, \mathbf{r}, \omega) & = - \hbar k_0^3 \left( n_\alpha (\omega) - n_\text{b} (\omega) \right) \sum_{l=1}^\infty \frac{2 l + 1}{8 \pi} \left( \mathcal{Q}^\text{E/H}_{l,\vartheta \varphi} \left[ \mathbf{e}_\vartheta \otimes \mathbf{e}_\vartheta + \mathbf{e}_\varphi \otimes \mathbf{e}_\varphi \right] + \mathcal{Q}^\text{E/H}_{l,r}  \mathbf{e}_r \otimes \mathbf{e}_r \right) , 
\label{eq:Q_EE_HH_sp} \\
\mathds{Q}_\text{EH} (\mathbf{r}, \mathbf{r}, \omega) & = \text{i} \hbar k_0^3 \left( n_\alpha (\omega) - n_\text{b} (\omega) \right) \sum_{l=1}^\infty \frac{2 l + 1}{8 \pi} \mathcal{Q}^\text{mix}_{l,\vartheta \varphi} \left[ \mathbf{e}_\vartheta \otimes \mathbf{e}_\varphi + \mathbf{e}_\varphi \otimes \mathbf{e}_\vartheta \right] .
\label{eq:Q_EH_sp}
\end{align}
Note that $\mathds{B}_{\text{HE}}^\dagger = \mathds{B}_{\text{EH}}$ and $\mathds{Q}_\text{HE}^\dagger = \mathds{Q}_\text{EH}$. The definitions of the $\mathcal{B}$ and $\mathcal{Q}$ expressions can be found in Eqs.~\eqref{eq:B_EH_tp}-\eqref{eq:Q_mix}. This results in the following final expressions for the mean values
\begin{align}
\Big \llangle u_\text{th}(\mathbf{r}, t) \Big \rrangle & = 2 \sum_{j \in \{r,\vartheta,\varphi\}} \sum_{\gamma \in \{\text{E,H}\}} \Bigr[ \Lambda^\text{eq}_{j,\gamma} (0) - \Lambda^\text{leq}_{j,\gamma} (0) \Bigr] = \varepsilon_0 \Big \llangle I(\mathbf{r}, t) \Big \rrangle 
\label{eq:u_sp}
\end{align}
and the variances
\begin{align}
\text{Var}_{u,\text{th}} (\mathbf{r}, \mathbf{r}, \tau) & = 8 \varepsilon_0^2 \bigg\{ \sum_{j \in \{r,\vartheta,\varphi\}} \sum_{\gamma \in \{\text{E,H}\}} \Bigl[\text{Re} \Bigl( \Lambda^\text{eq}_{j,\gamma} (\tau) - \Lambda^\text{leq}_{j,\gamma} (\tau) \Bigr) \Bigr]^2 + 2 \Bigl[ \text{Re} \Bigl( \Lambda^\text{eq}_\text{mix} (\tau) - \Lambda^\text{leq}_\text{mix} (\tau) \Bigr) \Bigr]^2 \notag \\
& \quad + 2 \Bigl[ \text{Re} \Bigl( \Lambda^\text{eq}_\text{mix} (-\tau) - \Lambda^\text{leq}_\text{mix} (-\tau) \Bigr) \Bigr]^2 \bigg\} , 
\label{eq:var_u_sp} \\
\text{Var}_I (\mathbf{r}, \mathbf{r}, \tau) & = 4 \bigg\{ \sum_{j \in \{r,\vartheta,\varphi\}} \sum_{\gamma \in \{\text{E,H}\}} \Big| \Lambda^\text{eq}_{j,\gamma} (\tau) - \Lambda^\text{leq}_{j,\gamma} (\tau) \Big|^2 + 2 \Big| \Lambda^\text{eq}_\text{mix} (\tau) - \Lambda^\text{leq}_\text{mix} (\tau) \Big|^2 \notag \\
& \quad + 2 \Big| \Lambda^\text{eq}_\text{mix} (-\tau) + \Lambda^\text{leq}_\text{mix} (-\tau) \Big|^2 \bigg\}
\label{eq:var_I_sp}
\end{align} 
where I defined the different $\Lambda$'s in Eqs.~\eqref{eq:L_eq_EH_rtp}-\eqref{eq:L_leq_mix}. Let me stress that it is, again, possible to rewrite the sum of the integrated matrices $\mathds{B}$ and $\mathds{Q}$ in terms of a diagonal matrix and a matrix containing their eigenvectors so that one is left with the sum over squared eigenvalues. This becomes obvious from Eq.~\eqref{eq:B_EE_HH_sp} and Eq.~\eqref{eq:Q_EE_HH_sp} since there are no tensor products of mixed unit vectors. By comparing the contributions for each index $j$ in Eqs.~\eqref{eq:u_sp}-\eqref{eq:var_I_sp}, it is also apparent that for the spherical geometry Eqs.~\eqref{eq:rel1_sum}-\eqref{eq:rel2_sum} will be fulfilled again. As one would expect, the expressions in Eqs.~\eqref{eq:u_sp}-\eqref{eq:var_I_sp} are independent of the angles $\vartheta$ and $\varphi$ which would obviously change if the two points $\mathbf{r}'$ and $\mathbf{r}$ are different.

Let me also consider the case of small spheres with respect to the radial distance meaning $R \ll r$. In this case I can restrict myself to only take into account the first multipole order $l=1$ which has been investigated in many studies, e.g.\ including the heat transfer between two spherical particles \cite{nara_2008}, between a sphere and a substrate \cite{Otey_2011}, and for a single sphere in general \cite{Chaumet_1998}. As a result, it is known to be a good approximation for distances $r$ of more than three times of the radius of the corresponding spherical particles \cite{Dong, Herz5}. Then, the transmission matrices in Eqs.~\eqref{eq:TM}-\eqref{eq:TN} reduce to 
\begin{align}
\mathcal{T}_1^M & = \text{i} \frac{2}{45} \varepsilon (k_0 R)^5, 
\label{eq:TM_l1} \\
\mathcal{T}_1^N & = \text{i} \frac{2 (\varepsilon - 1)}{3 (\varepsilon + 2)} (k_0 R)^3.
\label{eq:TN_l1}
\end{align} 
Since $k_0 R \ll 1$ holds, one can safely neglect the contribution of $T_1^M$ and $|\mathcal{T}_1^{M/N}|^2$. Then, Eqs.~\eqref{eq:u_sp}-\eqref{eq:var_I_sp} reduce to 
\begin{align}
\Big \llangle u_\text{th}(\mathbf{r}, t) \Big \rrangle & = 12 \Lambda_\text{vac} (\tau) + 2 \sum_{j \in \{r,\vartheta,\varphi\}} \sum_{\gamma \in \{\text{E,H}\}} \Lambda^{r \ll R}_{j,\gamma} (0) = \varepsilon_0 \Big \llangle I(\mathbf{r}, t) \Big \rrangle 
\label{eq:u_sp_lim}
\end{align}
and the variances
\begin{align}
\text{Var}_{u,\text{th}} (\mathbf{r}, \mathbf{r}, \tau) & = 8 \varepsilon_0^2 \bigg\{ \sum_{j \in \{r,\vartheta,\varphi\}} \sum_{\gamma \in \{\text{E,H}\}} \Bigl[\text{Re} \Bigl( \Lambda_\text{vac} (\tau) + \Lambda^{R \ll r}_{j,\gamma} (\tau) \Bigr) \Bigr]^2 + 2 \Bigl[ \text{Re} \Bigl( \Lambda^{R \ll r}_\text{mix} (\tau) \Bigr) \Bigr]^2 + 2 \Bigl[ \text{Re} \Bigl( \Lambda^{R \ll r}_\text{mix} (-\tau) \Bigr) \Bigr]^2 \bigg\} , 
\label{eq:var_u_sp_lim} \\
\text{Var}_I (\mathbf{r}, \mathbf{r}, \tau) & = 4 \bigg\{ \sum_{j \in \{r,\vartheta,\varphi\}} \sum_{\gamma \in \{\text{E,H}\}} \Big| \Lambda_\text{vac} (\tau) + \Lambda^{R \ll r}_{j,\gamma} (\tau) \Big|^2 + 2 \Big| \Lambda^{R \ll r}_\text{mix} (\tau) \Big|^2 + 2 \Big| \Lambda^{R \ll r}_\text{mix} (-\tau) \Big|^2 \bigg\} .
\label{eq:var_I_sp_lim}
\end{align} 
The corresponding abbreviations can be found in Eqs.~\eqref{eq:L_vac}-\eqref{eq:L_lim_mix}. Finally, I would like to investigate this limit in the far field for $r \rightarrow \infty$. In this case, the scattered contribution within the mean values and the variances always contains a product of the spherical Hankel function $h_1$ and spherical Bessel function $j_1$. Hence, an exponential function $e^{\text{i} k_0 r}$ will occur which diminishes the scattering contribution for large distances $r$ due to rapid oscillations. Therefore, the $\Lambda^{R \ll r}$ integrals reduce to the ones in Eqs.~\eqref{eq:L_ultlim_Er}-\eqref{eq:L_ultlim_mix}. When only taking into account the leading order in $r$, I end up with
\begin{align}
\Big \llangle u_\text{th}(\mathbf{r}, t) \Big \rrangle & = \frac{\pi^2 k_\text{B}^4 T_\text{b}^4}{15 c^3 \hbar^3} + \frac{3 \hbar R^3}{\pi^2 c^4 r^2} \int_0^\infty \mathrm{d} \omega \, \omega^4 \left( n_\alpha(\omega) - n_\text{b}(\omega) \right) \frac{\text{Im}(\varepsilon)}{|\varepsilon + 2|^2} = \varepsilon_0 \Big \llangle I(\mathbf{r}, t) \Big \rrangle
\label{eq:u_sp_ultlim}
\end{align} 
and
\begin{align}
\text{Var}_{u,\text{th}} (\mathbf{r}, \mathbf{r}, \tau) & = \frac{12 k_\text{B}^4 T_\text{b}^4}{\pi^4 c^6 \hbar^3} \bigg\{ \frac{k_\text{B}^4 T_\text{b}^4}{\hbar^3} \left[\text{Re} \left( \zeta \left( 4, 1 - \text{i} \frac{\tau}{\tau_\text{b}} \right) \right) \right]^2 \notag \\
& \quad + \frac{\hbar R^3}{c r^2} \text{Re} \left( \zeta \left( 4, 1 - \text{i} \frac{\tau}{\tau_\text{b}} \right)\right) \int_0^\infty \mathrm{d} \omega \, \omega^4 \left( n_\alpha(\omega) - n_\text{b}(\omega) \right) \frac{\text{Im}(\varepsilon)}{|\varepsilon + 2|^2} \cos(\omega \tau) \bigg\} 
\label{eq:var_u_sp_ultlim} \\
\text{Var}_I (\mathbf{r}, \mathbf{r}, \tau) & = \frac{6 k_\text{B}^4 T_\text{b}^4}{\pi^4 c^6 \hbar^3 \varepsilon_0^2} \bigg\{ \frac{k_\text{B}^4 T_\text{b}^4}{\hbar^3} \Big| \zeta \left( 4, 1 - \text{i} \frac{\tau}{\tau_\text{b}} \right) \Big|^2 \notag \\
& \quad + \frac{\hbar R^3}{c r^2} \text{Re} \left(\zeta \left( 4, 1 - \text{i} \frac{\tau}{\tau_\text{b}} \right)  \int_0^\infty \mathrm{d} \omega \, \omega^4 \left( n_\alpha(\omega) - n_\text{b}(\omega) \right) \frac{\text{Im}(\varepsilon)}{|\varepsilon + 2|^2} e^{-\text{i} \omega \tau} \right) \bigg\} .
\label{eq:var_I_sp_ultlim}
\end{align} 
Obviously, one will retrieve the vacuum result again either by setting $T_\alpha = T_\text{b}$ so that the sphere itself is not emitting radiation or by going in the extreme far field in which the second terms become negligibly small.

\begin{figure}[hbt]
    \center
    \includegraphics[width=0.7\textwidth]{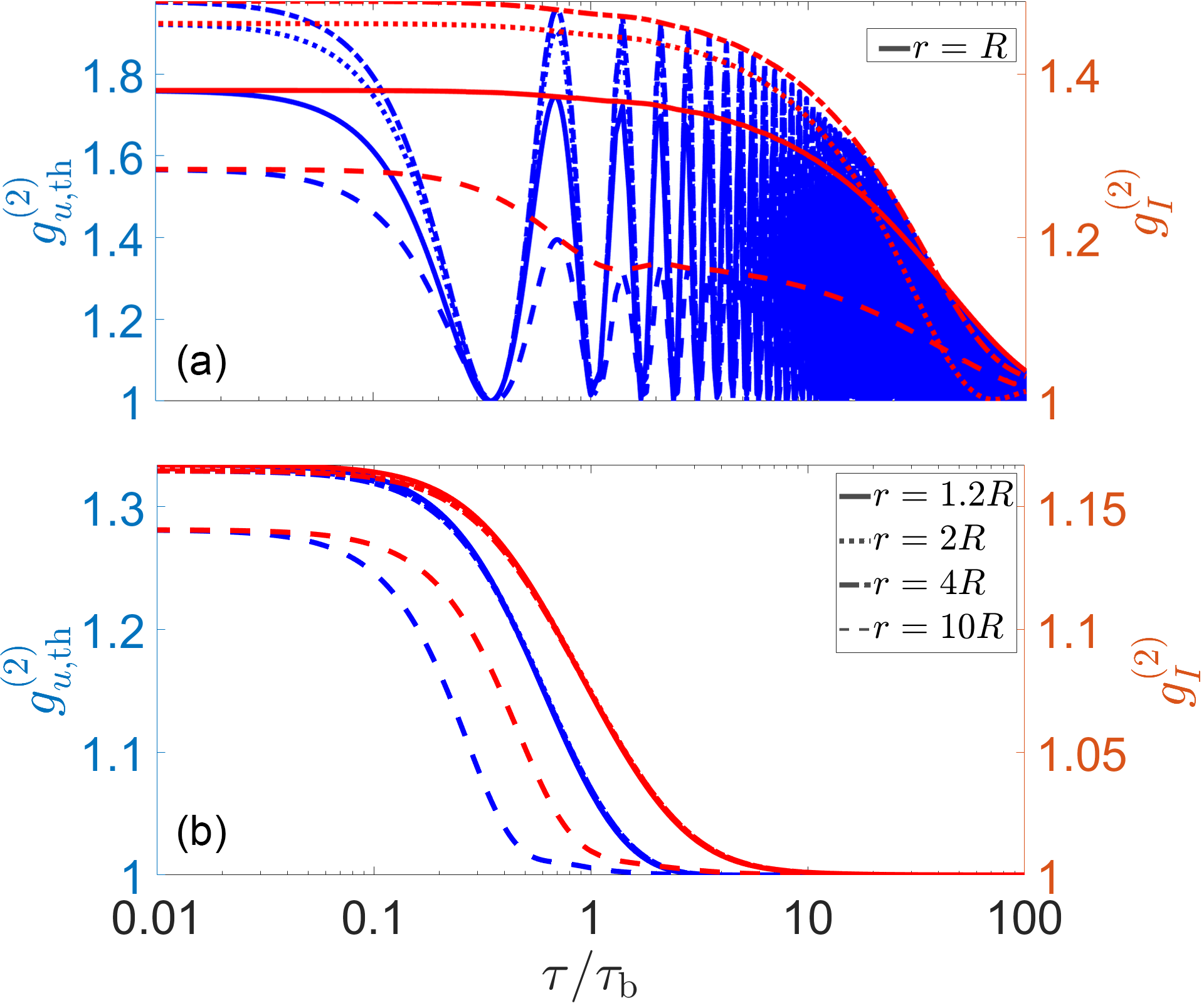}
    \caption{$g^{(2)}$ function of the thermal contribution of the energy density (blue) and intensity (red) for different distances $r$ between observation point and sphere's center for $T_\alpha = 700$ K and $T_\text{b} = 300$ K with respect to the normalized time delay $\tau/\tau_\text{b}$. The calculations are performed for the sphere's materials SiC (a) and gold (b) for $R=20$ nm.}
    \label{fig:t_sp}
\end{figure}
\begin{figure}[hbt]
    \center
    \includegraphics[width=0.7\textwidth]{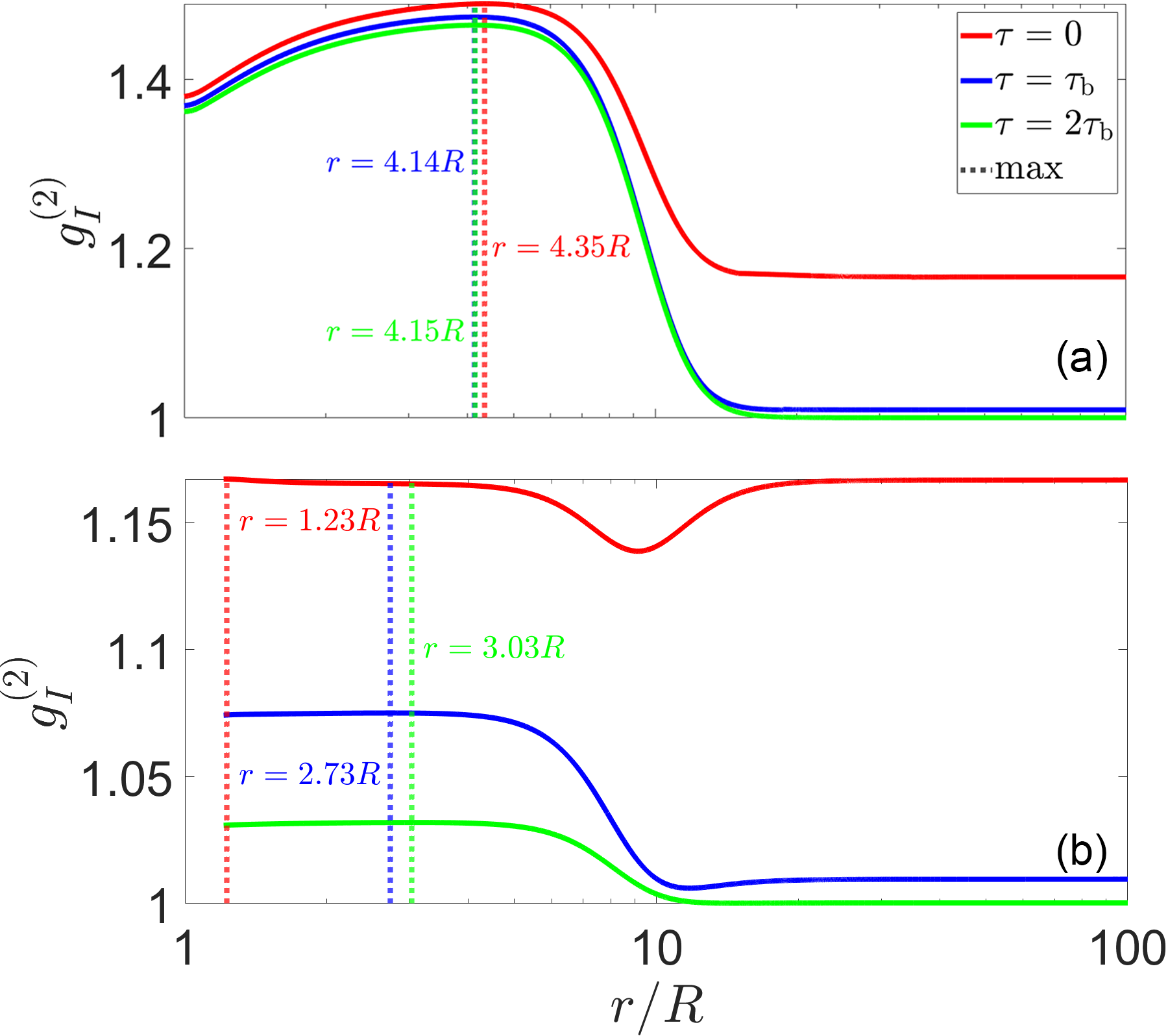}
    \caption{$g^{(2)}$ function of the intensity evaluated at different correlation times $\tau$ for $T_\alpha = 700$ K and $T_\text{b} = 300$ K with respect to the normalized distance $r/R$ for $R=20$ nm. The calculations are performed for the sphere's materials SiC (a) and gold (b). The maximal value of each curve is indicated by the vertical dashed lines.}
    \label{fig:r_sp}
\end{figure}

As an illustrating example, I show the functions $g_{u,\text{th}}^{(2)}$ (blue) and $g_{I}^{(2)}$ (red) for gold and SiC for different distances $r$ in Fig.~\ref{fig:t_sp}. The radius is $R=20$ nm and the temperatures are $T_\alpha = 700$ K and $T_\text{b} = 300$ K. The multipole order $l$ is chosen such that the difference to the values of the next order becomes insignificant. The actual meaning of that will be detailed out in the last figure of this section. Numerically, it was not possible to find such a value for gold for distances $r \sim R$ within the framework of the Drude model. This is a feature of the spherical Bessel functions $y_l = \text{Im} (h_l)$ becoming very large for small arguments and the Drude model which provides diverging values for small frequencies so that the numerical calculations strongly depend on the choice of the frequency range which was $5 \times 10^{12} \, \frac{\text{rad}}{\text{s}} < \omega < 10^{15} \, \frac{\text{rad}}{\text{s}}$. Therefore, reliable calculations for this model were only taken into account for distances up to $r = 1.2 R$. First of all, both $g^{(2)}$ functions basically show the same qualitative behavior regarding the dependence on time delay $\tau$ and radial distance $r$. The $g^{(2)}_{u, \text{th}}$ function for SiC shows an additional wave like character as mentioned in the previous chapter. For a SiC sphere, the localized surface phonon polariton (LSPhP) resonance frequency causes a quasi-monochromatic spectrum for the dominating electric part, whereas the TOP mode is, again, only present in the magnetic contribution. Gold does not have such resonances in the infrared regime and, therefore, $\tau_c^\text{gold} \approx \tau_\text{b}$ holds. For all curves I retrieve the bunching property of heat radiation but with a different distance behavior compared to the one for a half-space. SiC clearly shows that the photons are stronger correlated closer to the surface of the sphere, although the $g^{(2)}$ function finds a plateau between $3R \leq r \leq 5R$ (see Fig.~\ref{fig:r_sp}). However, the correlation time $\tau_c$ of smaller $r$ is now shorter than the one for larger $r$. For gold this maximum is close to the chosen minimal distance to the sphere's surface $r = 1.2 R$ and also only slightly larger than the one for the vacuum due to its broadband spectrum which is always approached for large distances $r$. 

To make this radial distance behavior more obvious, I show $g_{I}^{(2)}$ for gold and SiC in Fig.~\ref{fig:r_sp} for different $\tau$ depending on $r$. I also indicated the distance $r$ for which $g_{I}^{(2)}$ becomes maximal. Interestingly, the case $r=R$ is not the most likely one of finding bunched photons for SiC. This is rather the case at $r_{\rm max,SiC} \approx 4 R$. For gold this maximum is strongly dependent on $\tau$. Interestingly, at $\tau=0$ gold can also have values below vacuum at $2 R r < 11 R$ which has to be a geometrical effect because it only occurs in the far field. This is a feature of the local equilibrium contribution describing emission by the sphere itself because it vanishes for $T_\alpha = T_\text{b}$. Let me add that for larger temperature differences the distance $r$ for the most likely measurement of bunched photons also increases. Both materials reach the same value in the limit of large $r$ showing that for such distances the coherence properties correspond to that of the vacuum environment at $T_\text{b}=300$ K. Then, $g_{I}^{(2)}$ values are identical to those in section \ref{ch:ed_vac}. 

\begin{figure}[hbt]
    \center
    \includegraphics[width=0.7\textwidth]{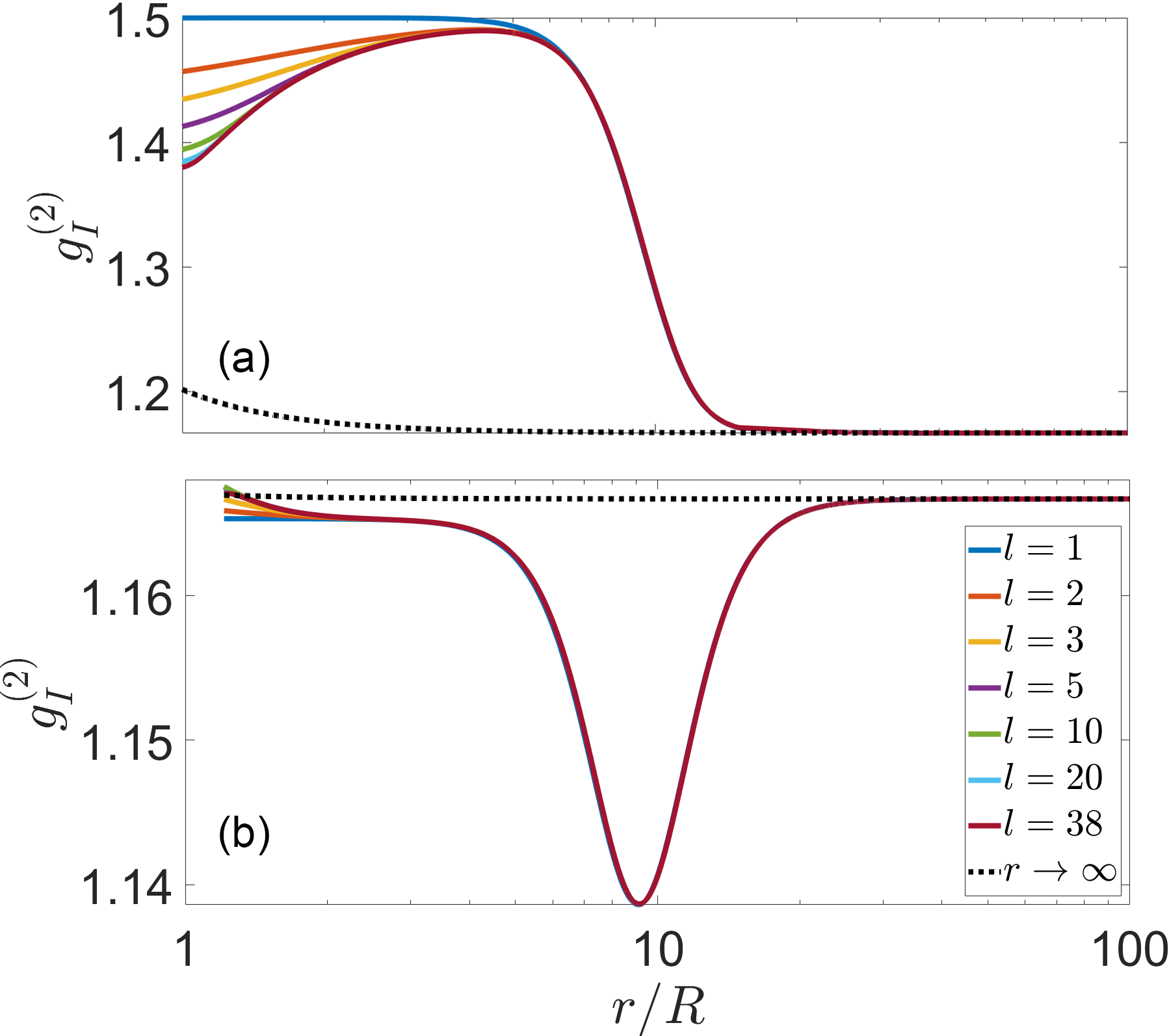}
    \caption{$g^{(2)}$ function of the intensity for SiC (a) and gold (b) evaluated at correlation time $\tau=0$ for $T_\alpha = 700$ K and $T_\text{b} = 300$ K with respect to the normalized distance $r/R$ for $R=20$ nm. The different lines correspond to different maximal multipole orders up to which the summation in Eq.~\eqref{eq:B_EE_HH_sp}-\eqref{eq:Q_EH_sp} is performed. Additionally the approximations in Eqs.~\eqref{eq:u_sp_ultlim}-\eqref{eq:var_I_sp_ultlim} are applied shown by the dashed lines.}
    \label{fig:l_sp}
\end{figure}
Eventually, let me come back to the evaluation with respect to the multipole orders. By that I refer to the highest multipole order used in the summations in Eq.~\eqref{eq:B_EE_HH_sp}-\eqref{eq:Q_EH_sp}. In Fig.~\ref{fig:l_sp} this is done for both materials, SiC and gold. There one can see that the dipole moment, e.g.\ $l=1$, dominates for distances starting at $r=5R$ for SiC and $r=2R$ for gold. This is in agreement with the approximation in Eqs.~\eqref{eq:u_sp_lim}-\eqref{eq:var_I_sp_lim} but already shows that one has to weigh carefully when to use it concerning the distance $r$ with respect to the material. Especially for distances $r<3R$, the dipole contribution strongly overestimates the overall result for SiC. For distances $r<1.5R$ even $l=5$ is insufficient to obtain an accurate result. For the smallest distance $r=R$ the multipole order $l=20$ can give accurate results. On the contrary, for gold an approximation of only using $l=1$ underestimates the overall result but for $r=1.2R$ the multipole order $l=20$ is sufficient. This also clearly shows that the multipole moments $l>1$ cause the above mentioned maximum value of $g_{I}^{(2)}$ at $3R<r<5R$ for SiC and an increase for $r \rightarrow R$ for gold. Therefore, in the above shown figures $l$ was always chosen such that the results can be regarded as exact for the considered distance $r$. In the cases of $r=R$ this means I chose $l=38$ for SiC.

\section{Extension to two objects}

To derive the corresponding expressions for $\Big \llangle u \Big \rrangle$, $\Big \llangle I \Big \rrangle$, $\text{Var}_{u,\text{th}}$, $\text{Var}_I$, $g^{(2)}_{u, \text{th}}$, and $g^{(2)}_I$ for two objects, one has to adapt the total fields and currents in Eqs.~\eqref{eq:J_1}-\eqref{eq:F_1} as it is explained in Ref.~\cite{Herz2}. The second object will be labeled by index $\beta$. Thereby, I obtain the fields
\begin{align}
\ket{\hat{\mathbf{F}}_k} & = \ket{\hat{\mathbf{F}}_{k, \text{b}}} + \text{i} \mu_0 \omega \mathds{G}_{k\text{E}} \left[ \ket{\hat{\mathbf{J}}_\alpha} + \ket{\hat{\mathbf{J}}_\beta} \right]
\end{align}
and the current density
\begin{align}
\ket{\hat{\mathbf{J}}_\alpha} & = \ket{\hat{\mathbf{J}}_{\alpha, \text{fl}}} + \frac{1}{\text{i} \mu_0 \omega} \mathds{T}_\alpha \ket{\hat{\mathbf{E}}_ \text{b}} + \mathds{T}_\alpha \mathds{G}_\text{EE} \Bigr[ \ket{\hat{\mathbf{J}}_{\beta, \text{fl}}} + \frac{1}{\text{i} \mu_0 \omega} \mathds{T}_\beta \ket{\hat{\mathbf{E}}_\text{b}} + \mathds{T}_\beta \mathds{G}_\text{EE} \ket{\hat{\mathbf{J}}_\alpha} \Bigr] .
\end{align}
Using this, I obtain the new correlation function
\begin{align}
\Big \llangle \ket{\hat{\mathbf{F}}_k} \otimes \bra{\hat{\mathbf{F}}_l} \Big \rrangle & = 2 \hbar \mu_0 \omega^2 \left[ \left[n_\text{b}(\omega) + 1 \right] \frac{\mathds{G}_{\text{f},kl} - \mathds{G}_{\text{f},lk}^\dagger}{2 \text{i}} + \sum_{\gamma \in \{\alpha,\beta\}} \left[ n_\gamma(\omega) - n_\text{b}(\omega) \right] \mathds{K}_\gamma \right]
\end{align}
with
\begin{align}
\mathds{G}_{\text{f}, kl} & = \left[\left(\mathds{1} + \mathds{O}_\alpha \mathds{G} \mathds{T}_\alpha + \mathds{O}_\beta \mathds{G} \mathds{T}_\beta \right) \mathds{G} \right]_{kl}, \\
\mathds{K}_{\gamma, kl} & = \left[ \mathds{O}_\gamma \mathds{G} \right]_{k \text{E}} \boldsymbol{\chi}_\gamma \left[ \mathds{G}^\dagger \mathds{O}_\gamma^\dagger \right]_{l \text{E}} , \\
\boldsymbol{\chi}_{\gamma, k} & = \frac{\mathds{T}_{\gamma, k} - \mathds{T}_{\gamma, k}^\dagger}{2 \text{i}} - \mathds{T}_{\gamma, k} \frac{\mathds{G}_{kk} - \mathds{G}_{kk}^\dagger}{2 \text{i}} \mathds{T}_{\gamma, k}^\dagger
\end{align}
as well as
\begin{align}
\mathds{O}_\alpha & = \left(\mathds{1} + \mathds{G} \mathds{T}_\beta \right) \left[ \mathds{1} - \mathds{G} \mathds{T}_\alpha \mathds{G} \mathds{T}_\beta \right]^{-1} .
\end{align}
The above equations, then, have to be inserted into Eqs.~\eqref{eq:Bkl}-\eqref{eq:Qkl} to get the mean values and variances for the energy density and intensity for two arbitrary objects in an arbitrary environment. This can also be extended to $N$ particles by regarding one of the two particles as a compound of $N-1$ particles, while repeating this procedure for that compound.

\section{Conclusion}

In this work I employed the methods of mQED and scattering approach to derive expressions for the variances of the energy density and intensity of heat radiation in a systems of an arbitrary object immersed in an arbitrary environment. I compared the general solution with the corresponding mean values and retrieved the expressions found in \cite{MandelWolf} for the ratio of variance and squared mean values for the intensity of isotropic systems. I also extended this to corresponding expressions of the energy density and systems containing three preferred axes like in cartesian, cylindrical, or spherical coordinates. With that formalism I computed the $g^{(2)}$ functions of both, the energy density and the intensity, for vacuum, a half-space, and a sphere. Thereby, I retrieved the results of Ref.~\cite{MandelWolf} for vacuum, showed the expected distance dependence of the $g^{(2)}$ functions above a half-space, and did the same thing for the $g_I^{(2)}$ of a sphere as well as a multipole order analysis. Interestingly, $g_I^{(2)}$ can become maximal for distances $r>R$ also depending on the material and the chosen temperatures, which seems to be due to higher order multipole moments. The effect behind this feature is unclear for the moment and might be interesting for future investigations. Finally, I showed theoretically how the expressions can be generalized for two arbitrary objects and a many-body system. Perhaps, experimental setups like in the HBT experiment can help to experimentally underpin the findings of this work.

\section{Acknowledgments}

The author acknowledges financial support by the Walter Benjamin Program of the Deutsche Forschungsgemeinschaft (eng. German Research Foundation) under project number 519479175 as well as fruitful discussions with PD Dr.\ Svend-Age Biehs (Carl von Ossietzky Universität Oldenburg, Germany).

%%%%%%%%%%%%%%%%%%%%%%%%%%%%%%%%%%%%%%%%%%%%%%%%%%%%%%%%%%%%%%%%%%%%%%%%%%%%%%%%%%%%%%%%%%%%%%%%%%%%%%%%%%%%%%%%%%%%%%%%%%%%%%%%%%%
%
%
%%%%%%%%%%%%%%%%%%%%%%%%%%%%%%%%%%%%%%%%%%%%%%%%%%%%%%%%%%%%%%%%%%%%%%%%%%%%%%%%%%%%%%%%%%%%%%%%%%%%%%%%%%%%%%%%%%%%%%%%%%%%%%%%%%%%%%
\appendix

\section{Green's functions and integral formulas for a planar geometry \label{appA}}

For the planar geometry, I use these expressions for the electric Green's function
\begin{align}
\mathds{G}_\text{vac,EE} (\mathbf{k}_\perp, z, z') & = \frac{\text{i} e^{\text{i} k_z (z-z')}}{2 k_z} \left[ \mathbf{a}_\perp (k_0) \otimes \mathbf{a}_\perp (k_0) + \mathbf{a}_\parallel^{+} (k_0) \otimes \mathbf{a}_\parallel^{+} (k_0) \right], \\
\mathds{G}_\text{scat,EE} (\mathbf{k}_\perp, z, z') & = \frac{\text{i} e^{\text{i} k_z (z+z')}}{2 k_z} \left[ r_\text{H} \mathbf{a}_\perp (k_0) \otimes \mathbf{a}_\perp (k_0) + r_\text{E} \mathbf{a}_\parallel^{+} (k_0) \otimes \mathbf{a}_\parallel^{-} (k_0) \right]
\label{eq:G_EE_sc}
\end{align}
and 
\begin{align}
\mathbf{k}_\perp & = (k_x, k_y)^T , 
\label{eq:kperp} \\
\mathbf{x}_\perp & = (x, y)^T , 
\label{eq:xperp} \\
\text{d}^2 k_\perp & = \text{d} k_x \text{d} k_y , 
\label{eq:d2kperp} \\
k_z & = \sqrt{k_0^2 - k_\perp^2} . 
\label{eq:kz}
\end{align}
$\mathds{G}_\text{vac}$ describes the vacuum part and $\mathds{G}_\text{scat}$ the contribution reflected at the substrate's surface. Here, I defined the polarization unit vectors
\begin{align}
\mathbf{a}_\perp (k_0) & = \frac{1}{k_\perp} (k_y, -k_x, 0)^T , 
\label{eq:aperp} \\
\mathbf{a}_\parallel^{\pm} (k_0) & = \frac{1}{k_\perp k_0} (\mp k_x k_z, \mp k_y k_z, k_\perp^2)^T,
\label{eq:apar}
\end{align}
and used the Fresnel amplitude reflection coefficients
\begin{align}
r_\text{s} & = \frac{k_z - k_{z,\text{sub}}}{k_z + k_{z,\text{sub}}} , 
\label{eq:rh} \\
r_\text{p} & = \frac{\varepsilon k_z - k_{z,\text{sub}}}{\varepsilon k_z + k_{z,\text{sub}}}
\label{eq:re}
\end{align}
with 
\begin{align}
k_{z,\text{sub}} & = \sqrt{\varepsilon k_0^2 - k_\perp^2}
\end{align}
where I introduced the substrates permittivity $\varepsilon$. 

For the mean values and variances I define the integrals 
\begin{align}
I_{\text{p/s}, \perp} & = \int_0^{k_0} \frac{\text{d} k_\perp k_\perp}{8 \pi k_z k_0} \text{Re} \left(e^{2 \text{i} k_z d} \left[ r_\text{s/p} - \frac{k_z^2}{k_0^2} r_\text{p/s} \right] \right) \notag \\
& \quad + \int_{k_0}^\infty \frac{\text{d} k_\perp k_\perp}{8 \pi |k_z| k_0} e^{- 2 |k_z| d} \text{Im} \left( r_\text{s/p} + \frac{|k_z|^2}{k_0^2} r_\text{p/s} \right) , 
\label{eq:I_ps_perp} \\
I_{\text{p/s}, z} & = \int_0^{k_0} \frac{\text{d} k_\perp k_\perp^3}{8 \pi k_z k_0^3} \text{Re} \left(e^{2 \text{i} k_z d} r_\text{p/s} \right) + \int_{k_0}^\infty \frac{\text{d} k_\perp k_\perp^3}{8 \pi |k_z| k_0^3} e^{- 2 |k_z| d} \text{Im} (r_\text{p/s}) , 
\label{eq:I_ps_z} \\
I_\text{mix} & = \int_0^{k_0} \frac{\text{d} k_\perp k_\perp}{16 \pi k_0^2} \text{Im} \biggl( e^{2 \text{i} k_z d} [r_\text{s}(\mathbf{k}_\perp) - r_\text{p}(\mathbf{k}_\perp)] \biggr) \notag \\
& \quad + \int_{k_0}^\infty \frac{\text{d} k_\perp k_\perp}{16 \pi k_0^2} e^{- 2 |k_z| d} \text{Im}[r_\text{s}(\mathbf{k}_\perp) - r_\text{p}(\mathbf{k}_\perp)]
\label{eq:I_mix} 
\end{align}
for the equilibrium contribution ($T_\alpha = T_{\rm b}$) and 
\begin{align}
K_{\text{p/s},\perp} & = \int_0^{k_0} \frac{\text{d} k_\perp k_\perp}{8 \pi k_z k_0} \left( [1 - |r_\text{s/p}(\mathbf{k}_\perp)|^2] + \frac{k_z^2}{k_0^2} [1 - |r_\text{p/s}(\mathbf{k}_\perp)|^2] \right) \notag \\
& \quad + \int_{k_0}^\infty \frac{\text{d} k_\perp k_\perp}{4 \pi |k_z| k_0} e^{- 2 |k_z| d} \left[ \text{Im}(r_\text{s}(\mathbf{k}_\perp)) + \frac{|k_z|^2}{k_0^2} \text{Im}(r_\text{p}(\mathbf{k}_\perp)) \right] , 
\label{eq:K_ps_perp} \\
K_{\text{p/s},z} & = \int_0^{k_0} \frac{\text{d} k_\perp k_\perp^3}{8 \pi k_z k_0^3} [1 - |r_\text{p/s}(\mathbf{k}_\perp)|^2] + \int_{k_0}^\infty \frac{\text{d} k_\perp k_\perp^3}{4 \pi |k_z| k_0^3} e^{- 2 |k_z| d} \text{Im}(r_\text{p}(\mathbf{k}_\perp)) , 
\label{eq:K_ps_z} \\
K^\text{pr}_\text{mix} & = \int_0^{k_0} \frac{\text{d} k_\perp k_\perp}{16 \pi k_0^2} [2 - |r_\text{s}(\mathbf{k}_\perp)|^2 - |r_\text{p}(\mathbf{k}_\perp)|^2]  , 
\label{eq:K_pr_mix} \\
K^\text{ev}_\text{mix} & = \int_{k_0}^\infty \frac{\text{d} k_\perp k_\perp}{8 \pi k_0^2} e^{- 2 |k_z| d} \left[\text{Im}(r_\text{p}(\mathbf{k}_\perp)) - \text{Im}(r_\text{s}(\mathbf{k}_\perp)) \right] 
\label{eq:K_ev_mix}
\end{align}
for the local equilibrium contribution ($T_\alpha \neq T_{\rm b}$). For the frequency integrals, I define the expressions
\begin{align}
\Gamma^\text{leq}_{\text{E/H},\perp/\parallel} (\tau) & = \frac{\hbar}{\varepsilon_0 \pi} \int_0^\infty \! \mathrm{d} \omega k_0^3 [n_\alpha (\omega) - n_\text{b} (\omega)] K_{\text{p/s},\perp/\parallel} e^{\text{i} \omega \tau} , 
\label{eq:G_leq_EH} \\
\Gamma^\text{leq}_\text{mix,pr/ev} (\tau) & = \frac{2 \hbar}{\varepsilon_0 \pi} \int_0^\infty \! \mathrm{d} \omega k_0^3 [n_\alpha (\omega) - n_\text{b} (\omega)] K^\text{pr/ev}_\text{mix} e^{\text{i} \omega \tau} , 
\label{eq:G_leq_mix} \\
\Gamma^\text{eq}_{\text{E/H},\perp/\parallel} (\tau) & = \frac{\hbar}{\varepsilon_0 \pi} \int_0^\infty \! \mathrm{d} \omega k_0^3 n_\text{b}(\omega) \left(\frac{1}{6 \pi} \left( 1 + \delta_{j \perp} \right) + I_{\text{p/s}, \perp/\parallel} \right) e^{\text{i} \omega \tau} , 
\label{eq:G_eq_EH} \\
\Gamma^\text{eq}_\text{mix} (\tau) & = \frac{2 \hbar}{\varepsilon_0 \pi} \int_0^\infty \! \mathrm{d} \omega k_0^3 n_\text{b}(\omega) I_\text{mix} e^{\text{i} \omega \tau} .
\label{eq:G_eq_mix}
\end{align}

In the limit of infinite distances $d \rightarrow \infty$, the $I$ and $K^\text{ev}_\text{mix}$ integrals vanish leaving me with
\begin{align}
K_{\text{p/s},\perp} & = \frac{1}{6 \pi} - \int_0^{k_0} \frac{\text{d} k_\perp k_\perp}{8 \pi k_z k_0} \left( |r_\text{s/p}(\mathbf{k}_\perp)|^2 + \frac{k_z^2}{k_0^2} |r_\text{p/s}(\mathbf{k}_\perp)|^2 \right) , 
\label{eq:K_ps_perp} \\
K_{\text{p/s},z} & = \frac{1}{12 \pi} - \int_0^{k_0} \frac{\text{d} k_\perp k_\perp^3}{8 \pi k_z k_0^3} |r_\text{p/s}(\mathbf{k}_\perp)|^2 , 
\label{eq:K_ps_z} \\
K^\text{pr}_\text{mix} & = \frac{1}{16 \pi} - \int_0^{k_0} \frac{\text{d} k_\perp k_\perp}{16 \pi k_0^2} [|r_\text{s}(\mathbf{k}_\perp)|^2 + |r_\text{p}(\mathbf{k}_\perp)|^2] .
\label{eq:K_pr_mix}
\end{align}
By using these integrals, I end up with the following $\Gamma$ integrals
\begin{align}
\Gamma^\text{eq}_{\text{E/H},\perp} (\tau) + \Gamma^\text{leq}_{\text{E},\perp} (\tau) & = \Gamma_{\text{vac},1} (\tau) - \Gamma^{d \rightarrow \infty}_{\text{E/H},\perp} (\tau) , 
\label{eq:G_EH_perp} \\
\Gamma^\text{eq}_{\text{E/H},\parallel} (\tau) + \Gamma^\text{leq}_{\text{E},\parallel} (\tau) & = \frac{1}{2} \Gamma_{\text{vac},1} (\tau) - \Gamma^{d \rightarrow \infty}_{\text{E/H},z} (\tau) ,
\label{eq:G_E_z} \\
\Gamma^\text{leq}_\text{mix,pr} (\tau) & = \Gamma_{\text{vac},2} (\tau) - \Gamma^{d \rightarrow \infty}_\text{mix,pr} (\tau) .
\label{eq:G_mix}
\end{align}
using the abbreviations
\begin{align}
\Gamma_{\text{vac},1} (\tau) & = \frac{k_\text{B}^4}{\varepsilon_0 \pi^2 c^3 \hbar^3} \left[ T_\alpha^4 \zeta \left(4, 1 - \text{i} \frac{\tau}{\tau_\alpha} \right) + T_\text{b}^4 \zeta \left(4, 1 - \text{i} \frac{\tau}{\tau_\text{b}} \right) \right] , 
\label{eq:G_vac1} \\
\Gamma_{\text{vac},2} (\tau) & = \frac{3 k_\text{B}^4}{4 \pi^2 \varepsilon_0 c^3 \hbar^3} \left[ T_\alpha^4 \zeta \left(4, 1 - \text{i} \frac{\tau}{\tau_\alpha} \right) - T_\text{b}^4 \zeta \left(4, 1 - \text{i} \frac{\tau}{\tau_\text{b}} \right) \right] , 
\label{eq:G_vac2} \\
\Gamma^{d \rightarrow \infty}_{\text{E/H},\perp} (\tau) & = \frac{\hbar}{8 \pi^2 \varepsilon_0} \int_0^\infty \! \mathrm{d} \omega k_0^2 [n_\alpha (\omega) - n_\text{b} (\omega)] \int_0^{k_0} \text{d} k_\perp \frac{k_\perp}{k_z} \left( |r_\text{s/p}(\mathbf{k}_\perp)|^2 + \frac{k_z^2}{k_0^2} |r_\text{p/s}(\mathbf{k}_\perp)|^2 \right) e^{\text{i} \omega \tau} , 
\label{eq:G_inf_EH_perp} \\
\Gamma^{d \rightarrow \infty}_{\text{E/H},z} (\tau) & = \frac{\hbar}{8 \pi^2 \varepsilon_0} \int_0^\infty \! \mathrm{d} \omega [n_\alpha (\omega) - n_\text{b} (\omega)] \int_0^{k_0} \text{d} k_\perp \frac{k_\perp^3}{k_z} |r_\text{p/s}(\mathbf{k}_\perp)|^2 e^{\text{i} \omega \tau} ,
\label{eq:G_inf_EH_z} \\
\Gamma^{d \rightarrow \infty}_\text{mix,pr} (\tau) & = \frac{\hbar}{8 \pi^2 \varepsilon_0} \int_0^\infty \! \mathrm{d} \omega k_0 [n_\alpha (\omega) - n_\text{b} (\omega)] \int_0^{k_0} \text{d} k_\perp k_\perp [|r_\text{s}(\mathbf{k}_\perp)|^2 + |r_\text{p}(\mathbf{k}_\perp)|^2] e^{\text{i} \omega \tau} .
\label{eq:G_inf_mix}
\end{align}

In the quasi-static limit only the evanescent contributions of all integrals will contribute to the overall result. For the $I$ and $K$ integrals I obtain
\begin{align}
K_{\text{p},\perp/z} & = \frac{1}{8 \pi k_0^3 d^3} \frac{\text{Im}(\varepsilon)}{|\varepsilon + 1|^2} = 2 I_{\text{p}, \perp/z}, 
\label{eq:K_qs_p} \\
K_{\text{s},\perp} & = \frac{1}{8 \pi k_0 d} \frac{\text{Im}(\varepsilon)}{|\varepsilon + 1|^2} = 2 I_{\text{s}, \perp} , 
\label{eq:K_qs_s} \\
K^\text{ev}_\text{mix} & = \frac{1}{16 \pi k_0^2 d^2} \frac{\text{Im}(\varepsilon)}{|\varepsilon + 1|^2} = -2 I_\text{mix}, 
\label{eq:K_qs_mix} \\
I_{\text{s}, z} & = K_{\text{s},z} = K^\text{pr}_\text{mix} = 0.
\label{eq:zeros}
\end{align}
This yields the following $\Gamma$ integrals
\begin{align}
\Gamma^\text{eq}_{\text{E},\perp/\parallel} (\tau) + \Gamma^\text{leq}_{\text{E},\perp/\parallel} (\tau) & = \frac{\hbar}{16 \pi^2 \varepsilon_0 d^3} \int_0^\infty \! \mathrm{d} \omega [2 n_\alpha (\omega) - n_\text{b} (\omega)] \frac{\text{Im}(\varepsilon)}{|\varepsilon + 1|^2} e^{\text{i} \omega \tau} , 
\label{eq:G_qs_E} \\
\Gamma^\text{eq}_{\text{H},\perp} (\tau) + \Gamma^\text{leq}_{\text{H},\perp} (\tau) & = \frac{\hbar}{16 \pi^2 \varepsilon_0 d} \int_0^\infty \! \mathrm{d} \omega k_0^2 [2 n_\alpha (\omega) - n_\text{b} (\omega)] \frac{\text{Im}(\varepsilon)}{|\varepsilon + 1|^2} e^{\text{i} \omega \tau} , 
\label{eq:G_leq_H} \\
\Gamma^\text{eq}_\text{mix} (\tau) - \Gamma^\text{leq}_\text{mix,ev} (\tau) & = - \frac{\hbar}{16 \pi^2 \varepsilon_0 d^2} \int_0^\infty \! \mathrm{d} \omega k_0 [2 n_\alpha (\omega) - n_\text{b} (\omega)] \frac{\text{Im}(\varepsilon)}{|\varepsilon + 1|^2} e^{\text{i} \omega \tau} , 
\label{eq:G_qs_mix}
\end{align}

\section{Green's functions and integral formulas for a spherical geometry \label{appB}}

The two general solutions of the electric field for the spherical geometry can be expressed by
\begin{align}
\mathbf{E}^\text{reg/out}_{\text{M},l,m} (\mathbf{r}) & = \sqrt{\frac{(-1)^m k_0}{l (l+1)}} \begin{Bmatrix} j_l (k_0 r) \\ h_l (k_0 r) \end{Bmatrix} \biggl[ \frac{1}{\sin(\vartheta)} \frac{\partial Y_l^m(\vartheta, \varphi)}{\partial \varphi} \mathbf{e}_\vartheta - \frac{\partial Y_l^m(\vartheta, \varphi)}{\partial \vartheta} \mathbf{e}_\varphi \biggr] , \\
\mathbf{E}^\text{reg/out}_{\text{N},l,m} (\mathbf{r}) & = \sqrt{\frac{(-1)^m k_0}{l (l+1)}} \frac{1}{k_0 r} \biggl[ l (l+1) \begin{Bmatrix} j_l (k_0 r) \\ h_l (k_0 r) \end{Bmatrix} Y_l^m(\vartheta, \varphi) \mathbf{e}_r \notag \\
& \quad + \frac{\partial}{\partial r} r \begin{Bmatrix} j_l (k_0 r) \\ h_l (k_0 r) \end{Bmatrix} \biggl( \frac{\partial Y_l^m(\vartheta, \varphi)}{\partial \vartheta} \mathbf{e}_\vartheta - \frac{1}{\sin(\vartheta)} \frac{\partial Y_l^m(\vartheta, \varphi)}{\partial \varphi} \mathbf{e}_\varphi \biggr) \biggr] .
\end{align}
The unit vectors $\mathbf{e}_{r/\vartheta/\varphi}$ either point into radial or angular directions of $\vartheta$ or $\varphi$. $j_l$ and $h_l$ denote the spherical Bessel function and the spherical Hankel function of the $l$th order, respectively. $Y_l^m$ denotes the spherical harmonics of order $m$ and degree $l$. For the matrices $\mathds{B}$ and $\mathds{Q}$ is used the abbreviations
\begin{align}
\mathcal{B}^\text{E/H}_{l,\vartheta \varphi} & = \mathcal{T}_l^{M/N} h_l(k_0 r) j_l (k_0 r) + \mathcal{T}_l^{N/M} \frac{1}{(k_0 r)^2} \frac{\partial (r h_l(k_0 r))}{\partial r} \frac{\partial (r j_l(k_0 r))}{\partial r}, 
\label{eq:B_EH_tp} \\
\mathcal{B}^\text{E/H}_{l,r} & = l (l+1) \mathcal{T}_l^{N/M} \frac{h_l(k_0 r) j_l(k_0 r)}{(k_0 r)^2} 
\label{eq:B_EH_r} 
\end{align}
as well as the mixed term
\begin{align}
\mathcal{B}^\text{mix}_{l,\vartheta \varphi} & = \frac{1}{k_0 r} \biggl[ \mathcal{T}_l^N j_l(k_0 r) \frac{\partial (r h_l(k_0 r))}{\partial r} + \mathcal{T}_l^M h_l(k_0 r) \frac{\partial (r j_l(k_0 r))}{\partial r} \notag \\
& \quad + \mathcal{T}_l^{M*} j_l(k_0 r) \frac{\partial (r h_l^{*}(k_0 r))}{\partial r} + \mathcal{T}_l^{N*} h_l^{*}(k_0 r) \frac{\partial (r j_l(k_0 r))}{\partial r} \biggr]
\label{eq:B_mix} 
\end{align}
and 
\begin{align}
\mathcal{Q}^\text{E/H}_{l,\vartheta \varphi} & = \left[ \text{Re} \left( \mathcal{T}_l^{M/N} \right) + |\mathcal{T}_l^{M/N}|^2 \right] |h_l(k_0 r)|^2 + \left[ \text{Re} \left( \mathcal{T}_l^{N/M} \right) + |\mathcal{T}_l^{N/M}|^2 \right] \frac{1}{(k_0 r)^2} \bigg| \frac{\partial (r h_l(k_0 r))}{\partial r} \bigg|^2, 
\label{eq:Q_EH_tp} \\
\mathcal{Q}^\text{E/H}_{l,r} & = l (l+1) \left[ \text{Re} \left( \mathcal{T}_l^{N/M} \right) + |\mathcal{T}_l^{N/M}|^2 \right] \frac{|h_l(k_0 r)|^2}{(k_0 r)^2} 
\label{eq:Q_EH_r} 
\end{align}
with the mixed term
\begin{align}
\mathcal{Q}^\text{mix}_{l,\vartheta \varphi} & = \frac{1}{k_0 r} \left( \left[ \text{Re} \left( \mathcal{T}_l^{M} \right) + |\mathcal{T}_l^{M}|^2 \right] h_l(k_0 r)  \frac{\partial (r h_l^{*}(k_0 r))}{\partial r} + \left[ \text{Re} \left( \mathcal{T}_l^{N} \right) + |\mathcal{T}_l^{N}|^2 \right] h_l^{*}(k_0 r) \frac{\partial (r h_l(k_0 r))}{\partial r} \right).
\label{eq:Q_mix}
\end{align}

For the spherical geometry I define the following frequency integral expressions
\begin{align}
\Lambda^\text{eq}_{j,\text{E/H}} (\tau) & = \frac{\hbar}{2 \pi \varepsilon_0} \int_0^\infty \mathrm{d} \omega k_0^3 n_\text{b}(\omega) \biggl[ \frac{1}{6 \pi} + (1 + \delta_{jr}) \sum_{l=1}^\infty \frac{2 l + 1}{4 \pi} \text{Re} \left(\mathcal{B}^\text{E/H}_{l,j} \right) \biggl] e^{\text{i} \omega \tau} , 
\label{eq:L_eq_EH_rtp} \\
\Lambda^\text{leq}_{j,\text{E/H}} (\tau) & = \frac{\hbar}{2 \pi \varepsilon_0} \int_0^\infty \mathrm{d} \omega k_0^3 \left( n_\alpha(\omega) - n_\text{b}(\omega) \right) \sum_{l=1}^\infty \frac{2 l + 1}{8 \pi} (1 + \delta_{jr}) \mathcal{Q}^\text{E/H}_{l,j} e^{\text{i} \omega \tau} , 
\label{eq:L_leq_EH_rtp}
\end{align}
and
\begin{align}
\Lambda^\text{eq}_\text{mix} (\tau) & = \frac{\hbar}{2 \pi \varepsilon_0} \int_0^\infty \mathrm{d} \omega k_0^3 n_\text{b}(\omega) \sum_{l=1}^\infty \frac{2 l + 1}{8 \pi} \text{Im} \left(\mathcal{B}^\text{mix}_l \right) e^{\text{i} \omega \tau} , 
\label{eq:L_eq_mix} \\
\Lambda^\text{leq}_\text{mix} (\tau) & = \frac{\hbar}{2 \pi \varepsilon_0} \int_0^\infty \mathrm{d} \omega k_0^3 \left( n_\alpha(\omega) - n_\text{b}(\omega) \right) \sum_{l=1}^\infty \frac{2 l + 1}{8 \pi} \mathcal{Q}^\text{mix}_l e^{\text{i} \omega \tau} .
\label{eq:L_leq_mix}
\end{align}
The T-operators are defined by 
\begin{align}
\mathcal{T}_l^M & = - \frac{j_l(y) \frac{\partial}{\partial x} [x j_l(x)] - j_l(x) \frac{\partial}{\partial y} [y j_l(y)]}{j_l(y) \frac{\partial}{\partial x} [x h_l(x)] - h_l(x) \frac{\partial}{\partial y} [y j_l(y)]}
\label{eq:TM}
\end{align}
and
\begin{align}
\mathcal{T}_l^N & = - \frac{\varepsilon j_l(y) \frac{\partial}{\partial x} [x j_l(x)] - j_l(x) \frac{\partial}{\partial y} [y j_l(y)]}{\varepsilon j_l(y) \frac{\partial}{\partial x} [x h_l(x)] - h_l(x) \frac{\partial}{\partial y} [y j_l(y)]}
\label{eq:TN}
\end{align}
with $x = k_0 R$ and $y = \sqrt{\varepsilon} x$ using the sphere's material's permittivity $\varepsilon$.

In the case of small spheres compared to the radial distance so that $R \ll r$ holds, only the $l=1$ component of each term in the $\mathds{B}$ and $\mathds{Q}$ matrices contribute to the overall result which become within this limit
\begin{align}
\mathcal{B}^\text{E}_{1,\vartheta \varphi} & = \mathcal{T}_1^{N} \frac{1}{(k_0 r)^2} \frac{\partial (r h_1(k_0 r))}{\partial r} \frac{\partial (r j_1(k_0 r))}{\partial r}, \\
\mathcal{B}^\text{H}_{1,\vartheta \varphi} & = \mathcal{T}_1^{N} h_1(k_0 r) j_1 (k_0 r)
\end{align}
and
\begin{align}
\mathcal{B}^\text{E/H}_{1,r} & = 2 \mathcal{T}_1^{N/M} \frac{h_1(k_0 r) j_1(k_0 r)}{(k_0 r)^2} , \\
\mathcal{B}^\text{mix}_{1,\vartheta \varphi} & = \frac{1}{k_0 r} \biggl[ \mathcal{T}_1^N j_1(k_0 r) \frac{\partial (r h_1(k_0 r))}{\partial r} + \mathcal{T}_1^{N*} h_1^{*}(k_0 r) \frac{\partial (r j_1(k_0 r))}{\partial r} \biggr] 
\end{align}
as well as
\begin{align}
\mathcal{Q}^\text{E}_{1,\vartheta \varphi} & = \text{Re} \left( \mathcal{T}_1^{N} \right)  \frac{1}{(k_0 r)^2} \bigg| \frac{\partial (r h_1(k_0 r))}{\partial r} \bigg|^2 , \\
\mathcal{Q}^\text{H}_{1,\vartheta \varphi} & = \text{Re} \left( \mathcal{T}_1^{N} \right) |h_1(k_0 r)|^2 
\end{align}
and
\begin{align}
\mathcal{Q}^\text{E/H}_{1,r} & = 2 \text{Re} \left( \mathcal{T}_l^{N/M} \right) \frac{|h_1(k_0 r)|^2}{(k_0 r)^2} \\
\mathcal{Q}^\text{mix}_{1,\vartheta \varphi} & = \frac{1}{k_0 r} \text{Re} \left( \mathcal{T}_1^{N} \right) h_1^{*}(k_0 r) \frac{\partial (r h_1(k_0 r))}{\partial r} .
\end{align}
This yields the following $\Lambda$ integrals
\begin{align}
\Lambda^\text{eq}_{r/\vartheta/\varphi,\text{E/H/mix}} (\tau) - \Lambda^\text{leq}_{r/\vartheta/\varphi,\text{E/H/mix}} (\tau) & = \Lambda_\text{vac} (\tau) + \Lambda^{R \ll r}_{r/\vartheta/\varphi,\text{E/H/mix}} (\tau)
\end{align}
while using
\begin{align}
\Lambda_\text{vac} (\tau) & = \frac{k_\text{B}^4 T_\text{b}^4}{2 \pi^2 c^3 \hbar^3 \varepsilon_0} \zeta \left( 4, 1 - \text{i} \frac{\tau}{\tau_\text{b}} \right).
\label{eq:L_vac}
\end{align}
together with 
\begin{align}
\Lambda^{R \ll r}_{r,\text{E/H}} (\tau) & = \frac{3 \hbar}{8 \pi^2 \varepsilon_0} \int_0^\infty \mathrm{d} \omega k_0^3 \biggl( 2 n_\text{b}(\omega) \text{Re} \left(\mathcal{B}^\text{E/H}_{1,r} \right) - \left( n_\alpha(\omega) - n_\text{b}(\omega) \right) \mathcal{Q}^\text{E/H}_{1,r} \biggr) e^{\text{i} \omega \tau} , \\
\Lambda^{R \ll r}_{\vartheta/\varphi,\text{E}} (\tau) & = \frac{3 \hbar}{16 \pi^2 \varepsilon_0} \int_0^\infty \mathrm{d} \omega k_0^3 \biggl( 2 n_\text{b}(\omega) \text{Re} \left(\mathcal{B}^\text{E}_{1,\vartheta \varphi} \right) - \left( n_\alpha(\omega) - n_\text{b}(\omega) \right) \mathcal{Q}^\text{E}_{1,\vartheta \varphi} \biggr) e^{\text{i} \omega \tau} , \\
\Lambda^{R \ll r}_{\vartheta/\varphi,\text{H}} (\tau)& = \frac{3 \hbar}{16 \pi^2 \varepsilon_0} \int_0^\infty \mathrm{d} \omega k_0^3 \biggl( 2 n_\text{b}(\omega) \text{Re} \left(\mathcal{B}^\text{H}_{1,\vartheta \varphi} \right) - \left( n_\alpha(\omega) - n_\text{b}(\omega) \right) \mathcal{Q}^\text{H}_{1,\vartheta \varphi} \biggr) e^{\text{i} \omega \tau}
\end{align}
and
\begin{align}
\Lambda^{R \ll r}_\text{mix} (\tau) & = \frac{3 \hbar}{16 \pi^2 \varepsilon_0} \int_0^\infty \mathrm{d} \omega k_0^3 \biggl( n_\text{b}(\omega) \text{Im} \left(\mathcal{B}^\text{mix}_1 \right) - \left( n_\alpha(\omega) - n_\text{b}(\omega) \right) \mathcal{Q}^\text{mix}_1 \biggr) e^{\text{i} \omega \tau} .
\label{eq:L_lim_mix}
\end{align}
In the far field limit $r \rightarrow \infty$ these integrals reduce to 
\begin{align}
\Lambda^{R \ll r}_{r,\text{E}} (\tau) & = \frac{3 \hbar R^3}{2 \pi^2 \varepsilon_0 c^2 r^4} \int_0^\infty \mathrm{d} \omega \, \omega^2 \left( n_\alpha(\omega) - n_\text{b}(\omega) \right) \frac{\text{Im}(\varepsilon)}{|\varepsilon + 2|^2} e^{\text{i} \omega \tau} , 
\label{eq:L_ultlim_Er} \\
\Lambda^{R \ll r}_{r,\text{H}} (\tau) & = \frac{\hbar R^5}{30 \pi^2 \varepsilon_0 c^4 r^4} \int_0^\infty \mathrm{d} \omega \, \omega^4 \left( n_\alpha(\omega) - n_\text{b}(\omega) \right) \text{Im} \left( \varepsilon \right) e^{\text{i} \omega \tau}
\label{eq:L_ultlim_Hr}
\end{align}
and
\begin{align}
\Lambda^{R \ll r}_{\vartheta/\varphi,\text{E/H}} (\tau) & = \frac{3 \hbar R^3}{8 \pi^2 \varepsilon_0 c^4 r^2} \int_0^\infty \mathrm{d} \omega \, \omega^4 \left( n_\alpha(\omega) - n_\text{b}(\omega) \right) \frac{\text{Im}(\varepsilon)}{|\varepsilon + 2|^2} e^{\text{i} \omega \tau} , 
\label{eq:L_ultlim_EHtp} \\
\Lambda^{R \ll r}_\text{mix} (\tau) & = \text{i} \frac{3 \hbar R^3}{8 \pi^2 \varepsilon_0 c^4 r^2} \int_0^\infty \mathrm{d} \omega \, \omega^4 \left( n_\alpha(\omega) - n_\text{b}(\omega) \right) \frac{\text{Im}(\varepsilon)}{|\varepsilon + 2|^2} e^{\text{i} \omega \tau} .
\label{eq:L_ultlim_mix}
\end{align}

\bibliographystyle{apsrev4-2}
\bibliography{Referenzen}

\end{document}